\documentclass[twocolumn,
prx
,superscriptaddress
,floatfix
,aps
,10pt
,showpacs]{revtex4-2}

\usepackage{graphicx}
\usepackage{amssymb}
\usepackage{amsmath}
\usepackage{multirow}
\usepackage{array,tabularx}
\usepackage{xcolor}
\usepackage{comment}

\usepackage[colorlinks,urlcolor=blue,citecolor=blue,linkcolor=blue]{hyperref}

\usepackage{textcomp}

\newcommand{\bra}[1]{\langle\left.{#1}\right|}
\newcommand{\ket}[1]{\left|{#1}\right.\rangle}

\newcommand{\iac}[1]{{\color{magenta}#1}}

\usepackage{float}
\usepackage{upgreek}
\usepackage{makeidx}
\DeclareGraphicsExtensions{.png,.jpg,.pdf,.mps,.gif,.bmp,.eps}
\graphicspath{{figures/}}
\usepackage{physics}

\begin{document}

\title{Anyonic molecules in atomic fractional quantum Hall liquids: a quantitative probe of fractional charge and anyonic statistics}

\author{A. Mu$\tilde{\mathrm{n}}$oz de las Heras}
\email{a.munozdelasheras@unitn.it}
\affiliation{INO-CNR BEC Center and Dipartimento di Fisica, Universit$\grave{a}$ di Trento, 38123 Trento, Italy}

\author{E. Macaluso}
\affiliation{INO-CNR BEC Center and Dipartimento di Fisica, Universit$\grave{a}$ di Trento, 38123 Trento, Italy}

\author{I. Carusotto}
\affiliation{INO-CNR BEC Center and Dipartimento di Fisica, Universit$\grave{a}$ di Trento, 38123 Trento, Italy}

\date{\today}

%
\begin{abstract}
We study the quantum dynamics of massive impurities embedded in a strongly interacting two-dimensional atomic gas driven into the fractional quantum Hall (FQH) regime under the effect of a synthetic magnetic field. For suitable values of the atom-impurity interaction strength, each impurity can capture one or more quasi-hole excitations of the FQH liquid, forming a bound molecular state with novel physical properties.
An effective Hamiltonian for such anyonic molecules is derived within the Born-Oppenheimer approximation, which provides renormalized values for their effective mass, charge and statistics by combining the finite mass of the impurity with the fractional charge and statistics of the quasi-holes. The renormalized mass and charge of a single molecule can be extracted from the cyclotron orbit that it describes as a free particle in a magnetic field. The anyonic statistics introduces a statistical phase between the direct and exchange scattering channels of a pair of indistinguishable colliding molecules, and can be measured from the angular position of the interference fringes in the differential scattering cross section. Implementations of such schemes beyond cold atomic gases are highlighted, in particular in photonic systems.
%
\end{abstract}

\maketitle

%
\section{Introduction}
\label{sec:Introduction}

The discovery of the fractional quantum Hall (FQH) effect in two-dimensional (2D) electron gases under a strong transverse magnetic field~\cite{Tsui_1982,Tong_2016,Goerbig} is a cornerstone of modern physics. Not only did it pave the way towards the study of topological phases of matter~\cite{Hasan_2010} but also changed the paradigm of the boson-fermion dicotomy when the possibility of observing quasi-particles with fractional statistics (and fractional charge) in 2D systems was proposed, the so-called \textit{anyons}~\cite{Leinaas_1977,Wilczek_1982,Halperin_1984,Arovas_1984,Wu_1984,Wu_1984_2}.
Such exotic quasi-particles have been predicted to arise as emergent excitations of FQH fluids with different properties depending on the fluid density and the applied magnetic field~\cite{Tong_2016}. In addition to their intrinsic interest as exotic quantum mechanical objects, in the recent years they have started attracting a lot of attention also for the crucial role that they are expected to play in the development of fault-tolerant quantum computers~\cite{Nayak_2008}.
The existence of fractionally charged quasi-particles on the edges of a 2D electron gas in the FQH state was confirmed by shot-noise experiments~\cite{DePicciotto_1997}. Individual fractionally-charged states localized in the bulk of a FQH have been imaged using a scanning single electron transistor in~\cite{martin2004localization}. In contrast, a clear signature of fractional statistics has long remained elusive~\cite{Willett_2009,Mourik_2012}. Very recently, the consequences of fractional statistics were observed in a mesoscopic electronic device as a generalized exclusion principle in a current-current correlation measurement at the output of a beamsplitter~\cite{Bartolomei_2020}.

In parallel to these studies in the electronic context of solid-state physics, impressive developments in the experimental study of  ultracold atomic gases~\cite{PitaevskiiStringariBook} opened the door to the exploration of topological phases of matter using these highly controllable quantum systems~\cite{Cooper_2019}. Several protocols have been investigated to drive a 2D gas of ultracold atoms into the FQH regime. Conceptually, the most straightforward one relies on the Coriolis force experienced by neutral atoms set into rotation, which formally recovers the Lorentz force felt by charged particles in a magnetic field. Alternative strategies to induce effective Lorentz forces on neutral atoms involve the application of suitable optical and magnetic fields in order to associate a non-trivial Berry phase to the atomic motion and generate a synthetic magnetic field~\cite{Dalibard_2011, Dalibard_2015}. At sufficiently low temperatures and for sufficiently strong interactions, the atoms are then expected to turn into a sequence of strongly-correlated FQH liquid states for growing values of the angular speed or of the synthetic magnetic field strength~\cite{Tong_2016,Goerbig,Cooper_AdvPhys.57.539}. Pioneering experimental investigations in this direction were reported in~\cite{Gemelke_Chu_arXiv:1007.2677}, whereas different protocols to access the FQH regime are being proposed for both bosonic and fermionic cold atomic gases~\cite{Popp_2004,Sorensen_2005,Regnault_2011,Grusdt_2014,Palm_2020,Andrade_2020}.

In the last decade, a new platform has emerged as a promising candidate to study many-body physics, including strongly correlated FQH liquids. Starting from microcavity polaritons in semiconductor microstructures, assemblies of photons in nonlinear optical devices are presently under active study as the so-called quantum fluids of light~\cite{Carusotto_2013,Carusotto_2020}. In addition to the effective mass (typically induced by the spatial confinement) and the binary interactions (mediated by the optical nonlinearity of the medium), recent developments have demonstrated synthetic magnetic fields and realized various topological models for light ~\cite{Ozawa_2019}. The first experimental study of the interplay of strong photon-photon interactions with a synthetic magnetic field was reported for a three-site system in~\cite{roushan2017chiral}. The realization of a two-particle Laughlin state was presented in~\cite{Clark_2020} using the giant nonlinearity of Rydberg polaritons and the synthetic magnetic field of a twisted optical cavity~\cite{Schine_2016}.

Simultaneously to these exciting experimental advances, theorists have started investigating new strategies to probe in an unambiguous way the anyonic nature of the excitations of quantum Hall fluids. A Ramsey-like interferometry scheme to detect the many-body braiding phase arising upon exchange of two anyons was proposed  for a cold atom cloud in~\cite{Paredes_2001}. A related proposal exploiting the peculiarities of driven-dissipative photonic systems was presented in~\cite{umucalilar2013many}. Spectroscopical consequences of the Haldane exclusion statistics were pointed out in~\cite{Cooper_2015} and soon translated to the photonic context in~\cite{Umucalilar_2017}. A subtle quantitative relation between the density profile of quasi-holes (QHs) and their anyonic statistics was theoretically put forward  in~\cite{Umucalilar_2018,Macaluso_2019} and numerically confirmed for discrete lattice geometries in~\cite{macaluso2020charge}. Finally, random unitary techniques to measure the many-body Chern number were investigated in~\cite{elben2019statistical}.

In analogy with polarons arising from the many-body dressing of an impurity immersed in a cloud of quantum degenerate atoms~\cite{Massignan_2014,Hu_2016,Jorgensen_2016}, a series of works~\cite{Zhang_2014,Zhang_2015,Lundholm_2016} have anticipated the possibility of using impurity particles immersed in a FQH liquid to capture quasi-hole excitations (that is, flux tubes) and thus generate new anyonic molecules that inherit the fractional statistics of the quasi-hole. Observable consequences of the fractional statistics were pointed out in the fractional angular momentum of the impurities and, correspondingly, in their correlation functions and density profiles~\cite{Zhang_2014,Grass_2020}. An interferometric scheme to measure fractional charges by binding a mobile impurity to quasi-particles was proposed in~\cite{grusdt2016interferometric}. Alternative models where heavy particles may acquire fractional statistics by interacting with phonons in the presence of strong magnetic fields and/or fast rotation were proposed in~\cite{Yakaboylu_2018,Yakaboylu_2019}. The transport properties of impurities embedded in a Fermi gas in a (integer) Chern-insulating state were recently studied in~\cite{Camacho-Guardian_2019}. Spectroscopic signatures of the fractional statistics were also anticipated for the threshold behaviour of the neutron scattering and particle tunneling cross sections of gapped quantum
spin liquids and fractional Chern insulators~\cite{Morampudi_17}.

In the present paper, we take inspiration from the aforementioned theoretical works and from the highly developed experimental techniques that are available to address and manipulate single atoms in large atomic gases to theoretically illustrate how such anyonic molecules are a very promising tool to observe fractional statistics and shine new light on the microscopic physics of FQH fluids.
In particular, we investigate the quantum mechanical motion of a few anyonic molecules. Capitalizing on previous works, we provide a rigorous derivation of the effective Hamiltonian starting from a controlled Born-Oppenheimer (BO) approximation~\cite{BO,Scherrer_2017} where the positions of the impurities play the role of the slow degrees of freedom and the surrounding FQH fluid provides the fast ones. Whereas bare quasi-holes typically do not support motional degrees of freedom~\footnote{In the simplest example of a Laughlin state, quasi-holes have a vanishing energy that does not depend on position and do not possess any intrinsic kinetic energy. In a trapped configuration, the position of a quasi-hole evolves in response to the trapping potential according to a first-order differential equation. This dynamics is analogous to the one of vortices in superfluids~\cite{Fetter}.}, the anyonic molecule is found to display a fully fledged spatial dynamics, with a mass determined by the impurity mass supplemented by a non-trivial correction due to the quasi-hole inertia. 
Binding to the QH also modifies the effective charge of the impurity by including the Berry phase~\cite{Resta_2000} that the QH accumulates during its motion in space. All together, an anyonic molecule then behaves as a free charged particle in a magnetic field, whose cyclotron radius provides detailed information on the renormalized mass and on the fractional charge.

In the presence of two anyonic molecules, the fractional statistics of the QHs results in a long-range Aharonov-Bohm-like interaction between them. We illustrate the consequences of this long-range topological interaction in the simplest scattering process where two such objects are made to collide. For both hard-disk and dipolar interaction potentials, we calculate the differential scattering cross section for indistinguishable impurities, finding that for large relative momenta it features alternate maxima and minima due to the interference of direct and exchange scattering channels: analogously to textbook two-slit experiments, the interference pattern rigidly shifts when the statistical phase that the anyonic molecules acquire upon exchange is varied. This interference effect is instead suppressed when distinguishable impurities are considered. Experiments along these lines would therefore allow to confirm the existence of particles beyond the traditional boson-fermion classification and to quantitatively measure the statistics of the QHs in a direct way.

The structure of the article is the following. In Sec.~\ref{sec:Model} we review the system Hamiltonian and in Sec.~\ref{sec:BO} we develop the rigorous Born-Oppenheimer framework that we employ to study the quantum dynamics of the anyonic molecules: In Subsec.~\ref{sec:SingleImpurity} we establish the single particle parameters of the anyonic molecule and in Subsec.~\ref{sec:TwoImpurities} we recover the interaction Hamiltonian between molecules. The theory of two-body scattering is presented in Sec.~\ref{sec:TwoImpuritiesAnyPolScatt}, where we summarize our predictions for the angular dependence of the differential scattering cross-section and we highlight the qualitative impact of the fractional statistics. 
Conclusions are finally drawn in Sec.~\ref{sec:Conclusions}.

%
\section{The physical system and the model}
\label{sec:Model}
%

We consider a system of quantum particles confined to the two-dimensional $x$-$y$ plane and formed by a small number $N$ of mobile impurities of mass $M$ immersed in a large bath of $n\gg N$ atoms of mass $m$ in a FQH state. For simplicity, in what follows the former will be indicated as {\em impurities}, while the latter will be indicated as {\em atoms}. A transverse and spatially uniform synthetic magnetic field $\bold{B}=B\, \bold{u}_{\text{z}}$ is applied to the whole system (where $\bold{u}_{\text{z}}$ is the unit vector in the $z$ direction), and we consider that the impurities and the atoms possess effective (synthetic) charges $Q$ and $q$, respectively. 

In the particular case in which the magnetic field is generated by rotating the trap around the $z$ axis, the value of these quantities is set by the atomic masses and the rotation frequency of the trap $\omega_{\text{rot}}$ via $qB=2m\omega_{\rm rot}$ and $QB=2M\omega_{\rm rot}$~\cite{Cooper_2008, Dalibard_2015}. In the spirit of Ref.~\cite{Gemelke_2010}, the corresponding centrifugal force can be compensated by harmonic trap potentials acting on each atomic species. Their strength has to be adjusted to give the same trapping frequency $\omega_{\text{hc}}=\omega_{\text{rot}}$ for the two species.

Even though our discussion will be focused on atomic systems, all our conclusions directly extend to any other platform where quantum particles are made to experience a synthetic gauge field and strong interparticle interactions, for instance photons in twisted cavity set-ups where Landau levels~\cite{Schine_2016} and Laughlin states~\cite{Clark_2020} have been recently observed. In this case Rydberg atoms of two different species giving rise to strongly interacting Rydberg polaritons will play the role of the atoms and the impurities. Keeping this correspondence in mind,
the model presented in this section, as well as the experiments proposed in Secs.~\ref{sec:SingleImpurity} and~\ref{sec:TwoImpuritiesAnyPolScatt}, can be directly translated to the optical platform.

In $\hbar=1$ units, the system Hamiltonian then reads
\begin{align}
    H=T_{\text{a}}+T_{\text{i}}+V_{\text{aa}}+V_{\text{ia}}+V_{\text{ii}} \;,
\end{align}
where
\begin{align}
    &T_{\text{a}}(\{\bold{r}_{j}\}) = \sum_{j=1}^{n} \frac{1}{2m}\left[-i\bold{\nabla}_{\bold{r}_{j}}-q\bold{A}(\bold{r}_{j})\right]^2 \; ,\\
    &T_{\text{i}}(\{\bold{R}_{j}\}) = \sum_{j=1}^{N} \frac{1}{2M}\left[-i\bold{\nabla}_{\bold{R}_{j}}-Q\bold{A}(\bold{R}_{j})\right]^2 \; ,\\
    &V_{\text{aa}}(\{\bold{r}_{j}\}) = g_{\text{aa}}\sum_{i<j}^{n}
    \delta(\bold{r}_{i}-\bold{r}_{j}) \; ,\\
    &V_{\text{ia}}(\{\bold{r}_{j}\},\{\bold{R}_{j}\}) = \sum_{i=1}^{n}\sum_{j=1}^{N} v_{\text{ia}}(\bold{r}_{i}-\bold{R}_{j}) \; ,\\
    &V_{\text{ii}}(\{\bold{R}_{j}\}) = \sum_{i<j}^{N} v_{\text{ii}}(\bold{R}_{i}-\bold{R}_{j}) \; .
\end{align}
We denote by $\bold{r}_{j}$ and $-i\bold{\nabla}_{{\bold{r}}_{j}}$ the position and canonical momentum of the $j$-th atom, while $\bold{R}_{j}$ and $-i\bold{\nabla}_{{\bold{R}}_{j}}$ represent those of the $j$-th impurity.
$\bold{A}(\bold{r}_{j})=B(-y_{j}/2,x_{j}/2,0)$ and $\bold{A}(\bold{R}_{j})=B(-Y_{j}/2,X_{j}/2, 0)$ are the vector potentials corresponding to the synthetic magnetic field ($\bold{B}=\bold{\nabla}\cross\bold{A}$) at the positions of atoms and impurities, respectively.

The strength of the contact binary interaction between atoms is quantified by the $g_{\text{aa}}$ parameter~\footnote{Choosing a contact interaction potential, we are implicitly focusing on bosonic atomic fluids. Extension to fermionic fluids is straightforward and just requires using a finite-range form of the interaction potential $v_{\rm aa}$ to stabilize the FQH state. All our conclusions hold in this case as well.}, whereas $v_{\text{ia}}$ and $v_{\text{ii}}$ denote the impurity-atom and impurity-impurity interaction potentials, respectively. 
When the synthetic magnetic field is large enough, the number of vortices $n_{\text{v}}$ in the atomic fluid becomes comparable with the number $n$ of atoms. At low enough temperatures and for sufficiently strong repulsive atom-atom interactions $g_{\text{aa}}$, the atomic gas enters the so-called FQH regime described by a rational value of the filling fraction $\nu=n/n_{\text{v}}$~\cite{Tong_2016,Goerbig,Cooper_2008}. This incompressible state is characterized by excitations with fractional charge and statistics (quasi-holes and quasi-particles).

As it was first anticipated in~\cite{Zhang_2014, Zhang_2015, Lundholm_2016} a repulsive interaction potential $v_{\text{ia}}$ between the impurities and the atoms leads to the pinning of quasi-hole excitations at the impurities' positions. As a result, quasi-holes adiabatically follow the motion of the impurity forming composite objects that can be regarded as {\em anyonic molecules}.
By looking at the density pattern of quasi-hole excitations shown in~\cite{Macaluso_2017,Macaluso_2018}, we anticipate that the number of quasi-holes pinned by each impurity can be controlled via the  strength of $v_{\text{ia}}$: a stronger and/or longer-ranged interaction will provide space for more quasi-holes bound to each impurity. For the sake of simplicity, in this work we will focus on the case of a single quasi-hole per impurity but generalization to the many quasi-holes case is straightforward.

As a final assumption, we will focus on impurity-impurity potentials $v_{\text{ii}}$ of a far larger range than both the atom-atom and impurity-atom interactions and the QH extension. This will allow us to work with impurities that are separated enough in space to give independent and non-overlapping anyonic molecules that interact via the $v_{\text{ii}}$ potential with no correction due to the microscopic structure of the quasi-holes. In particular we will focus on interaction potentials with hard-disk or dipolar spatial shapes.

%
\section{The Born-Oppenheimer approximation}
\label{sec:BO}
Several authors have theoretically addressed the quantum mechanics of mobile impurities immersed in FQH fluids and have written effective Hamiltonians for the motion of the resulting charge--flux-tube complexes~\cite{Zhang_2014,Zhang_2015,Lundholm_2016,grusdt2016interferometric,Camacho-Guardian_2019,Yakaboylu_2018,Yakaboylu_2019}. Most such treatments were however based on heuristic models of the binding mechanism: while this was sufficient to get an accurate answer for the synthetic charge and the fractional statistics, it did not provide a quantitative prediction for the mass of the anyonic molecule: this is in fact determined by the bare mass of the impurity, supplemented by a correction due to the inertia of the FQH quasi-holes.

To fill this gap, in this Section we will summarize a rigorous approach to this problem. The reader that is already familiar with such effective Hamiltonians and is not interested in the technical details and in the quantitative value of the parameters can jump to the experimental remarks in the final Subsec.~\ref{subsec:Expt} and then move on to the scattering theory in Sec.~\ref{sec:TwoImpuritiesAnyPolScatt}.

%
\subsection{General framework}

Our theoretical description is based on a Born-Oppenheimer formalism in which we treat the impurities' positions as the slowly-varying degrees of freedom, while those of the surrounding atoms play the role of the fast ones~\cite{BO,Mead92,Resta_2000}. For each position of the impurities, the atoms are assumed to be in their many-body ground state, which contains quasi-holes at the impurities' positions to minimize the repulsive interaction energy. Given the spatial coincidence of the impurity and the quasi-hole, in the following  the positions of the resulting molecules will be indicated with the same variables $\mathbf{R}_{i}$.
While our approach is known to be exact for fixed impurities, it extends to moving impurities as long as their kinetic energy is smaller than the energy gap between the quasi-hole state and its first excited state.

Under this approximation, the total wave function can be factorized as 
\begin{align}
\label{eq:ExactFactorization}
    \psi(\{\bold{r}_{i}\},\{\bold{R}_{i}\},t)=\varphi^{(0)}_{\{\bold{R}_{i}\}} (\{\bold{r}_{i}\}) \, \chi(\{\bold{R}_{i}\},t) \;,
\end{align}
where the wave function $\chi(\{\bold{R}_{i}\},t)$ describes the quantum motion of the impurities and the atomic wave function 
$\varphi^{(0)}_{\{\bold{R}_{i}\}}(\{\bold{r}_{i}\})$
is the ground state of the Born-Oppenheimer atomic Hamiltonian
\begin{align}
    H_{\text{BO}}=T_{\text{a}}+V_{\text{aa}}+V_{\text{ia}}
\label{eq:BO_0th_H}
\end{align}
that includes the kinetic and interaction energy of the atoms and the interaction potential between atoms and impurities.
In what follows, we will use the shorthands $\bold{r}$ and $\bold{R}$ to denote the sets of atom coordinates $\{\bold{r}_{i}\}$ and of impurity coordinates $\{\bold{R}_{i}\}$.

In our specific FQH case with $\nu=1/w$ with positive and integer-valued $w$, the atomic wave function can be written 
in terms of the magnetic length $\ell_{\text{B}}=1/\sqrt{qB}$ and the complex in-plane coordinates $z=x-iy$ of the atoms as a many-quasi-hole wave function of the Laughlin form
\begin{align}
\label{eq:LaughlinStateQH}
     \varphi^{(0)}_{\bold{R}}(\bold{r})=\frac{1}{\sqrt{\mathcal{N}}}\prod_{i=1}^{n}\prod_{j=1}^{N}(z_{i}-Z_{j})\phi_{\text{L}}(\{z_{i}\}).
\end{align}
The positions of the quasi-holes are parameterically fixed by the (complex) positions $Z=X-iY$ of the impurities, while the last factor $\phi_{\text{L}}$ is the well-known Laughlin wave function of the FQH state~\cite{Laughlin_1983},
\begin{align}
\label{eq:LaughlinState}
    \phi_{\text{L}}(\{z_{i}\})=\prod_{i<j}^{n}(z_{i}-z_{j})^{1/\nu} e^{-\sum_{i=1}^{n}|z_{i}|^2/4\ell_{\text{B}}^2} \; .
\end{align}
In Eq.~\eqref{eq:LaughlinStateQH}, the normalization constant $\mathcal{N}$ is chosen to ensure the partial normalization condition
\begin{align}
\label{eq:PartialNormalization}
    \int d^{2n}\bold{r}\,|\varphi^{(0)}_{\bold{R}}(\bold{r})|^2=1 \; .
\end{align}
Provided that the impurities live in the bulk of the atomic cloud far from its edges and from each other, the energy $\epsilon^{(0)}_{\text{BO}}$ of the Born-Oppenheimer ground state is independent of the impurities' positions 
$\bold{R}$
and can be safely neglected.

The dynamics of the anyonic molecules will be governed by an effective Hamiltonian acting on the molecule wave function 
$\chi(\mathbf{R})$,
\begin{align}
\label{eq:EffHamDefinition}
H_{\text{eff}} =\bra{\varphi^{(0)}
_{\mathbf{R}}
}H\ket{\varphi^{(0)}
_{\mathbf{R}}
} 
\; ,
\end{align}
which, as we will discuss in full detail in the following subsections, takes a form
\begin{align}
\label{eq:EffHamGeneral}
H_{\text{eff}}=\sum_{j=1}^{N}\frac{\left[-i\bold{\nabla}_{\bold{R}_{j}}-\mathcal{Q}\bold{A}(\bold{R}_{j})+\bold{\mathcal{A}}_{\text{stat},j}
(\bold{R})\right]^2}{2\mathcal{M}}+V_{\text{ii}}(\bold{R})
\end{align}
that combines the properties of impurities and quasi-holes. 

Within this picture, each molecule then features a mass $\mathcal{M}$ --only approximately equal to the one of the impurities, see Sec.~\ref{sec:SingleImpurity_mass}-- and a total charge $\mathcal{Q}=Q-\nu q$ resulting from the sum of the bare charge $Q$ of the impurity and the one $-\nu q$ of the quasi-hole -- see Sec.~\ref{sec:SingleImpurity_charge}. These values are of course only accurate as long as the impurities are located in a region of constant density of the atomic cloud, that is, apart from each other in the bulk of an incompressible FQH phase. Under this condition, both the BO energy resulting from the interaction with the atoms $\epsilon^{(0)}_{\text{BO}}$ and the scalar potential arising in the BO approximation give spatially constant energy shifts that can be safely neglected.

In addition to these single-particle properties, the molecules inherit the interaction potential 
$V_{\text{ii}}(\bold{R})$ 
between the impurities and
experience a Berry connection $\bold{\mathcal{A}}_{\text{stat},j}(\bold{R})$ that now depends on the position of all molecules and encodes their quantum statistics.
For the Abelian FQH states under investigation here, we will see in Sec.~\ref{sec:TwoImpurities} that the effect of the Berry connection $\bold{\mathcal{A}}_{\text{stat},j}$ can be summarized by a single statistical parameter determined by the filling $\nu$ of the FQH atomic fluid, 
which indicates that the statistical phase picked upon exchange of two molecules is $\exp(i\pi\nu)$. If more $N_{\text{qh}}>1$ quasi-holes were pinned to the same impurity, the statistical phase would grow quadratically as  $N_{\text{qh}}^2\nu$~\cite{Tong_2016}.


Finally, note that we are restricting our attention to anyonic molecules that are separated enough in space for their internal structure not to be distorted by the interactions with the neighboring molecules. This is expected to be an accurate approximation if the inter-impurity distance is much larger than the range of the atom-impurity potential and the internal size of the quasi-hole --typically of the order of the magnetic length $\ell_{\text{B}}$~\cite{Tong_2016}. Under this approximation, the values of the renormalized mass and of the synthetic charge that we obtain for single molecules directly translate to the many-molecule case, and the interaction potential reduces to the inter-impurity one $v_{\rm ii}$ with no corrections from the microscopic structure of the molecules.

%
\subsection{Effective Hamiltonian for a single anyonic molecule}
\label{sec:SingleImpurity}

In this subsection we will investigate the parameters in the effective Hamiltonian (\ref{eq:EffHamGeneral}) that control the single-particle physics of the molecules, namely the renormalized mass $\mathcal{M}$ and charge $\mathcal{Q}$. A simple experimental configuration to extract these values will also be proposed at the end of the subsection.

%
\subsubsection{Mass renormalization}
\label{sec:SingleImpurity_mass}

A crucial, yet often disregarded feature of the BO approximation is the renormalization of the effective mass of the slow degrees of freedom. In molecular physics such a renormalization affects the effective mass of the nuclei dressed by the electrons and is essential to guarantee consistency of the description~\cite{Scherrer_2017,Resta_2000}. In our case it concerns the change of the effective mass of the impurity when this is dressed by the quasi-hole excitation in the surrounding FQH fluid. As far as we know, this feature was always overlooked in previous literature, even though it may give a quantitatively significant bias to observable quantities such as the effective magnetic length considered in~\cite{Zhang_2014}.

In order to obtain a quantitative estimate for the effective mass $\mathcal{M}$, we generalize the molecular physics approach of Ref.~\cite{Scherrer_2017} by including the synthetic magnetic field in the formalism. As it is discussed in detail in Appendix~\ref{sec:AppA}, one needs to include the first perturbative correction to the BO adiabatic approximation, which amounts to taking into account the distortion of the quasi-hole profile due to the motion of the impurity.
To this purpose we expand $\varphi_{\bold{R}}(\bold{r},t)\simeq \varphi^{(0)}_{\bold{R}}(\bold{r})+\varphi^{(1)}_{\bold{R}}(\bold{r},t)$,
where the BO wave function $\varphi^{(0)}_{\bold{R}}(\bold{r})$ is obtained as the ground state of the Hamiltonian (\ref{eq:BO_0th_H}) in the presence of a single impurity at $\bold{R}$ and has the quasi-hole form \eqref{eq:LaughlinStateQH} with $N=1$. While $\varphi^{(0)}_{\bold{R}}(\bold{r})$ only depends on the coordinate difference $\mathbf{r}-\mathbf{R}$, the first order perturbative correction $\varphi^{(1)}_{\bold{R}}(\bold{r},t)$ depends on the impurity speed. 

Following the theory of Ref.~\cite{Scherrer_2017}, the mass tensor of the molecule is then given at first order by
\begin{align}
\label{eq:MassRenormalization}
    \underline{\underline{\mathcal{M}}}=\underline{\underline{M}}+\Delta\underline{\underline{M}} \; ,
\end{align}
where $\underline{\underline{M}}$ is the $2\cross 2$ unity matrix multiplied by the bare impurity mass and the correction is such that the corresponding kinetic energy
%
\begin{align}
    \frac{1}{2} \Delta{{{M}}}_{\alpha\beta} v_\alpha v_\beta=\int d\bold{r}\,{\varphi^{(1)*}_{\bold{R}}(\bold{r},t)}\left[H_{\text{BO}}-\epsilon_{\text{BO}}^{(0)}(\bold{R})\right]{\varphi^{(1)}_{\bold{R}}(\bold{r},t)}
\label{eq:DeltaM}
\end{align}
recovers the increase in the BO energy due to the motion of the impurity. 

The correction ${\varphi^{(1)}_{\bold{R}}(\bold{r},t)}$ 
to the atomic wave function is obtained at the lowest perturbative level in the impurity speed $\underline v=(v_{\text{X}},v_{\text{Y}})$ by applying the inverse of the fast Hamiltonian
\begin{align}
    \left[H_{\text{BO}}-\epsilon^{(0)}_{\text{BO}}\right]{\varphi^{(1)}_{\bold{R}}(\bold{r},t)}=v_\alpha \, {\bold{\nabla}_\alpha}\varphi^{(0)}_{\bold{R}}(\bold{r})
\label{eq:phi1}
\end{align}
to the gradient of the atomic wave function with respect to the in-plane coordinates of the impurity $\alpha=\{X,Y\}$. 
In physical terms, this correction is such that the action of the fast BO Hamiltonian $H_{\text{BO}}$ recovers the temporal evolution of the BO wave function $\varphi_{\bold{R}(t)}(\bold{r})$ due to the spatial displacement of the impurity. 

In our Laughlin case, it is easy to show that the gradient of the atomic wave function with respect to $\bold{R}$ is proportional to the wavefunction of the lowest excited state of $H_{\text{BO}}$,
\begin{align}
    \nabla_{\bold{R}}\varphi^{(0)}_{\bold{R}}(\bold{r})=\frac{\tau}{\ell_{B}}\varphi^{(\text{e})}_{\bold{R}}(\bold{r}) \; . \label{eq:tau}
\end{align}
This excited state corresponds to a chiral $\Delta L=-1$ oscillation of the quasi-hole around the impurity~\footnote{In the case of many quasi-holes bound to a strong impurity potential, such an excitation mode can be viewed as the $\Delta L=-1$ chiral mode of the inner edge of the ring-shaped FQH fluid surrounding the impurity~\cite{Macaluso_2018}.} and, for the simplest case of a single impurity located at $Z=0$, its wavefunction (normalized as in \eqref{eq:PartialNormalization}) has the form
\begin{equation}
\varphi^{(\text{e})}_{\bold{R}}(\bold{r})=\frac{1}{\sqrt{\mathcal{N^{(\text{e})}}}}\sum_{i_o=1}^{n}\prod_{\substack{i=1 \\ i\neq i_o}}^n z_i\,\phi_{\text{L}}(\{z_i\})\,. 
\end{equation}
While these results are enough to establish that the correction to the mass tensor is diagonal in the $X,Y$ coordinates, a quantitative estimation of its magnitude needs microscopic insight on the overlap numerical factor $\tau$ and on the energy $\Delta\omega_{-1}$ of the excited state under consideration. 

This requires microscopic information on the atomic wavefunctions and a few technical steps, that are reported in the Appendices. It turns out from the calculations in Appendix~\ref{sec:AppB} that the numerical factor $\tau$ is of order one, in particular $\tau\sim 0.7$.
As we discuss in the Appendix~\ref{sec:AppC}, the excitation energy $\Delta\omega_{-1}$ grows for stronger impurity-atom interaction potentials $V_{\text{ia}}$ which reinforce the internal rigidity of the composite quasi-hole-impurity. Quantitatively, the excitation energy 
$\Delta\omega_{-1}$ 
is at most on the order of a fraction of the bulk many-body gap above the fractional quantum Hall state, namely a fraction of the atom-atom interaction energy scale $V_0=g_{\text{aa}}/2\ell_{\text{B}}^2$~\cite{Macaluso_2018}. Note also that the many-body gap can not exceed the cyclotron energy associated to the synthetic magnetic field acting on the atoms, $\omega_{\rm cycl}=1/m\ell_B^2= qB/m$.

These results can be plugged into Eq.~\eqref{eq:DeltaM}, giving a mass correction
\begin{align}
\label{eq:phi_e}
    \Delta M =\frac{2\tau^{2}}{
    \Delta \omega_{-1} \ell_{\text{B}}^2} \; ,
\end{align}
%
which, since $2\tau^2\simeq 1$, provides the figure of merit
\begin{align}
    \frac{\Delta M}{M}\simeq \frac{m}{M}\frac{\omega_{\text{cycl}}}{\Delta\omega_{-1}}
    \label{eq:DeltaMrelative}
\end{align}
that can be used to quantify its relative importance.

For instance, a small $m/M$ ratio can be obtained in a cold atoms experiment embedding heavy atoms like Erbium as impurities in a gas of light atoms such as Lithium. This already gives $m/M\simeq 0.04$ and can be further decreased replacing the heavy atom with a multi-atom molecule~\cite{Carr_2009}. On the other hand, the $\Delta\omega_{-1}/\omega_{\rm cycl}$ factor is typically bound to values below unity, but can be maximized using strong (perhaps Feshbach-enhanced~\cite{Chin_2010}) interactions among atoms and between atoms and impurities, so to push the many-body gap towards the cyclotron energy. All together, it is natural to expect that the correction to the effective impurity mass may be sizable and important to interpret the experimental observations.

%
\subsubsection{Synthetic charge}
\label{sec:SingleImpurity_charge}

As in the previous subsection, we consider the simplest situation in which a single quasi-hole is bound to a single impurity located at position $\bold{R}$. In this case the effective Hamiltonian~\eqref{eq:EffHamDefinition} takes the form
\begin{align}
    H_{\text{eff}}=\frac{\left[-i\bold{\nabla}_{\bold{R}}-Q\bold{A}(\bold{R})+\bold{\mathcal{A}}(\bold{R})\right]^2}{2\mathcal{M}}
    +\Phi(\bold{R})+\epsilon^{(0)}_{\text{BO}}(\bold{R})\;,
\end{align}
where
\begin{equation}
    \mathcal{A}(\bold{R})=-i\bra{\varphi^{(0)}_{\bold{R}}(\bold{r})}\nabla_{\bold{R}}\ket{\varphi^{(0)}_{\bold{R}}(\bold{r})}
\end{equation}
is the Berry connection related to the quasi-hole motion across the FQH fluid, which enters the equation above in the form of an effective vector potential~\cite{Mead92,Resta_2000}.

For the Laughlin states under consideration here, the Berry connection $\bold{\mathcal{A}}(\bold{R})$ can be calculated making use of the plasma analogy as reviewed in~\cite{Tong_2016}. This gives
\begin{align}
\bold{\mathcal{A}}(\bold{R})=\frac{q \nu}{2\ell_{\text{B}}^2}\bold{u}_{\text{z}}\times \bold{R}=-\nu q \bold{A}(\bold{R}) \; ,
\end{align}
%
which means that the QH feels the synthetic magnetic field as a fractional charge $-\nu q$ corresponding to the atomic density that has been displaced away from its surroundings. As a result, the effective single-molecule Hamiltonian can be recast in a compact form 
\begin{equation}
\label{eq:Heff_single}
    H_{\text{eff}} 
    =\frac{\left[-i\bold{\nabla}_{\bold{R}}-\left(Q-\nu q\right)\bold{A}(\bold{R})\right]^2}{2\mathcal{M}}
\end{equation}
in terms of an effective charge
\begin{equation}
    \mathcal{Q}=Q-\nu q
    \label{eq:Qrenorm}
\end{equation}
resulting from the sum of the bare charge $Q$ of the impurity and the fractional charge $-\nu q$ of the quasi-hole that is bound to it.

The effective scalar potential is instead equal to  
\begin{equation}
    \Phi(\bold{R})=\frac{-1}{2\mathcal{M}}\left[\bra{\varphi^{(0)}_{\bold{R}}(\bold{r})}\bold{\nabla}^2_{\bold{R}}\ket{\varphi^{(0)}_{\bold{R}}(\bold{r})}+\bold{\mathcal{A}}^2(\bold{R})\right]\; :
    \label{eq:scalarPhi}
\end{equation}
as long as the impurity lives in the bulk of the (incompressible) FQH fluid where the fluid density is --to a high precision-- constant, both the BO energy $\epsilon^{(0)}_{\text{B0}}$ and the scalar potential are constant and can be safely neglected.

In order to facilitate the description of the two-body scattering process and isolate the features of interest, it will be beneficial to design an experiment where the effective charge $\mathcal{Q}$ of the anyonic molecules vanishes. From Eq.~\eqref{eq:Qrenorm} one sees that if the gauge field is generated by rotation the $\mathcal{Q}=0$ condition translates in a ratio $M/m=\nu$ between the masses of both atomic species, which is not straightforwardly compatible with the assumptions underlying the BO approximation. On the other hand, a careful design of the optical and magnetic fields applied to the atoms allows to tune the strength of the synthetic magnetic field acting on each of them, so to satisfy the required $\mathcal{Q}=0$ condition.

As an alternative strategy, even though the $\mathcal{Q}=0$ condition is not naturally fulfilled in the laboratory reference frame, the effect of the finite $\mathcal{Q}$ can be removed by looking at the system from  a reference frame that rotates around the $z$ axis at an angular frequency $\tilde{\Omega}$  such that 
\begin{equation}
\tilde{\Omega}=-\frac{\mathcal{Q}\,\mathbf{B}}{2\mathcal{M}}.
\end{equation}
Under this condition, the Coriolis force associated to the rotation is equal and opposite to the effective synthetic Lorentz force acting on the anyonic molecule. Inserting a value of the impurity mass on the order of the one of a heavy (e.g. Erbium) atom and a synthetic magnetic field on the order of $\mathcal{Q}\mathbf{B}\sim 1/\lambda^2$ with $\lambda$ in the optical range $\lambda\sim 1\,\mu\textrm{m}$, one finds a value for $\tilde{\Omega}$ in an accessible $100\,\textrm{Hz}$ range.

But one must not forget that moving to the rotating frame not only introduces a Coriolis force, but also transforms the velocity appearing in the synthetic magnetic Lorentz force, which results in an additional force to be added to usual centrifugal force of rotating reference frames. In the system under consideration here, all such centrifugal/centripetal forces can be  compensated in the rotating frame by introducing an additional anti-harmonic trapping potential $V_{\rm i}(\mathbf{R})=-\frac{1}{2}\mathcal{M}\tilde{\Omega}^2 \mathbf{R}^2$
acting on each impurity. Combining all different terms, the dynamics of isolated anyonic molecules in the rotating reference frame is then the desired one of free particles moving along straight lines~\footnote{The need for an unusual anti-trapping potential can be physically understood since we wish to transform localized cyclotron orbits into open straight lines going to infinity.} for which the scattering process will be the simplest. 

Before concluding, it is worth stressing that the rotation at $\tilde{\Omega}$ considered here is just a way of looking at the effective dynamics of the anyonic molecules, and does not affect the underlying atoms that form a FQH state in the laboratory frame in the presence of the synthetic magnetic field~\footnote{The situation is of course slightly more complex if a rotating atomic gas is  used to generate the FQH state. In this case the two rotations and the two contributions to the centrifugal potential must be carefully combined. Note also that, thanks to the $R_{\rm rel}^{-2}$ dependence of the two-body term in the Berry connection \eqref{eq:Berry_two_terms}, the rotation at $\tilde{\Omega}$ has no effect on this latter term. Writing the Lagrangian associated to the Hamiltonian \eqref{eq:Ham_two_body}, transformation of the two-body term to the rotating frame only provides an additional constant term.}.

\begin{figure}
    \centering
    \includegraphics[width=0.5\textwidth]{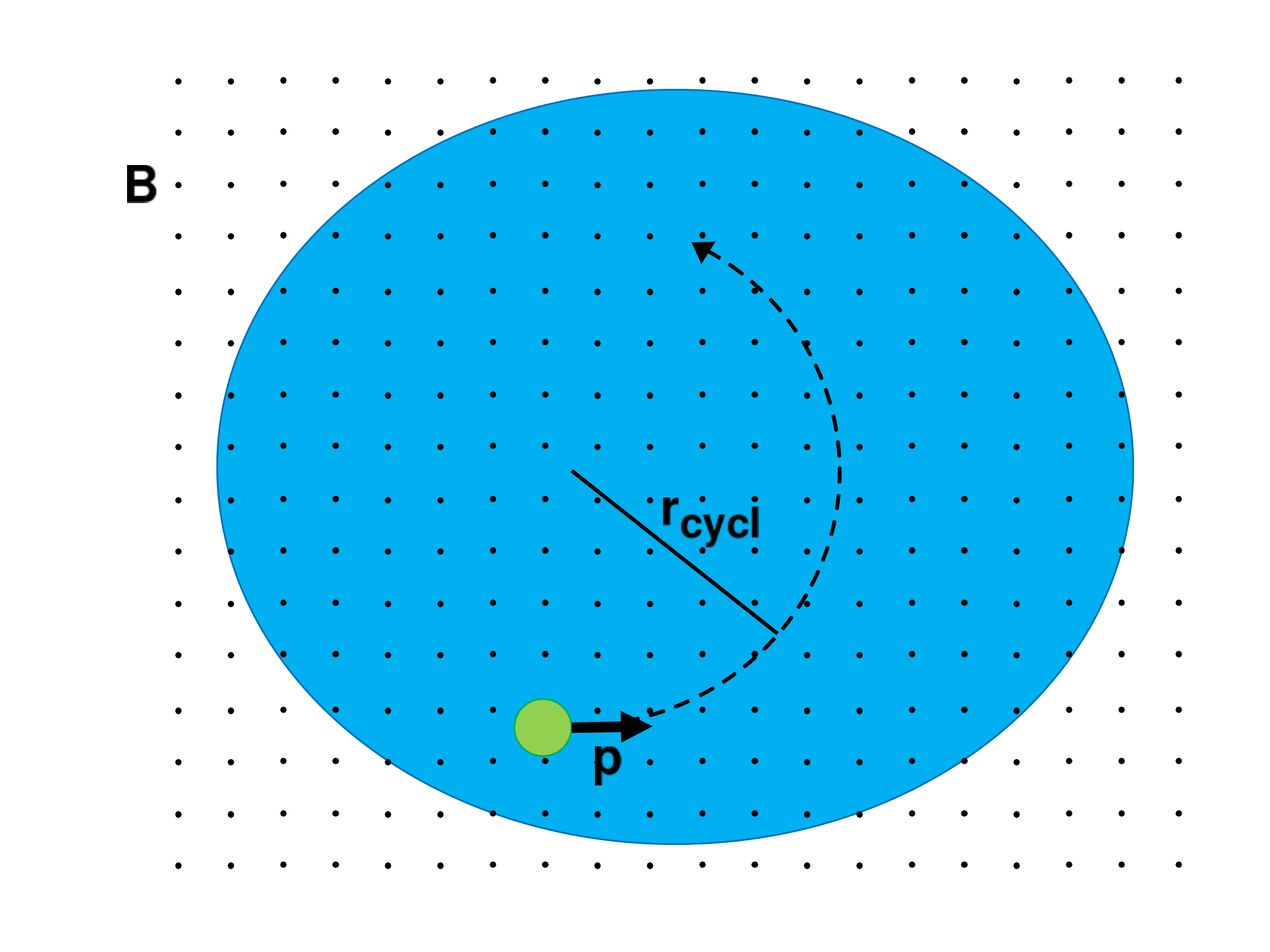}
    \caption{{\em Proposed experiment to measure the renormalized mass and the fractional charge of an anyonic molecule.} An impurity atom (green circle) is located in the bulk of a fractional quantum Hall fluid of atoms (blue region). Its repulsive interactions with the atoms make it bind one or more FQH quasi-holes, forming a composite {\em anyonic molecule} with renormalized mass $\mathcal{M}$, fractional (synthetic) charge $\mathcal{Q}$, and anyonic fractional statistics.
    To measure the renormalized mass and charge, one can impart a momentum kick $\bold{p}$ to a single such molecule initially at rest and follow the ensuing cyclotron motion. 
    }
    \label{fig:ExpSingleImp}
\end{figure}

\subsubsection{Experimental remarks}
\label{subsec:Expt}

Since the Hamiltonian \eqref{eq:Heff_single} describes a free particle in a magnetic field, we can envisage a simple experiment to measure the fractional charge and the renormalized mass of the molecule from the radius of its cyclotron orbit, as sketched in Fig.~\ref{fig:ExpSingleImp}. Once the molecule receives a momentum kick $\bold{p}$ (e.g. by applying a time-dependent force to the impurity), it starts describing a cyclotron orbit. For a given value of the momentum kick, the molecule mass $\mathcal{M}$ can be directly obtained from the actual speed via $\bold{p}=\mathcal{M}\bold{v}$. The charge is then extracted from the cyclotron radius via the textbook formula
\begin{equation}
r_{\text{cycl}}=\frac{\mathcal{M} v}{\mathcal{Q} B}.
\end{equation}
In order to determine the fractional charge, it is useful to consider the ratio $\mathcal{Q}/Q$ which is obtained by comparing the cyclotron radius for a molecule immersed in the FQH fluid with the one of the bare impurity in the absence of the surrounding FQH fluid. The relation \eqref{eq:Qrenorm} allows then to relate the observed charge $\mathcal{Q}$ to the fractional charge of the quasi-holes in the FQH fluid. If the synthetic magnetic field is generated by rotating the system, its calibration is made even simpler by the fact that the product $qB$ ($QB$) is determined by the rotation speed $\omega_{\text{rot}}$ via $qB=2m\omega_{\text{rot}}$ ($QB=2M\omega_{\text{rot}}$)~\cite{Dalibard_2015,Cooper_AdvPhys.57.539}. 

Taking advantage of the different nature of the impurity particle as compared to the atoms forming the FQH fluid, reconstruction of the trajectory of the anyonic molecule can be done by imaging the position of the impurity at different evolution times after its deterministic preparation at a given location with a known momentum imparted, e.g. by an external potential. In this respect, we can expect that using impurities with a large mass offers the further advantage of a more accurate definition of the initial position and velocity against Heisenberg indetermination principle. 

This issue can be put in quantitative terms by comparing the maximum kinetic energy of the molecule compatible with the BO approximation, with the cyclotron energy of the molecule in the synthetic magnetic field which quantifies its zero-point motion: in order for the cyclotron motion to be visible, several Landau levels must in fact be populated. The maximum kinetic energy can be estimated from $K_{\rm max}=\mathcal{M} v_{\rm max}^2/2$ using the velocity $v_{\rm max}$ at which the norm of the first-order correction $\varphi_{\mathbf{R}}^{(1)}$ in \eqref{eq:phi1} becomes of order one, namely
\begin{equation}
    v_{\rm max}=\frac{\Delta\omega_{-1}\,\ell_B}{\tau}\,.
\end{equation}
The cyclotron energy is given by the usual formula using the effective charge $\mathcal{Q}$ of the molecule, namely $\Omega_{\rm cycl}=\mathcal{Q} B/\mathcal{M}$. Imposing $K_{\rm max}\gg \Omega_{\rm cycl}$ then requires
\begin{equation}
    \sqrt{\frac{2\tau^2 \mathcal{Q}}{q}}\frac{m}{\mathcal{M}}\frac{\omega_{\rm cycl}}{\Delta\omega_{-1}} \ll 1
\end{equation}
which, recalling that $2\tau^2\simeq 1$, is related to the condition \eqref{eq:DeltaMrelative} for a small mass correction and is well satisfied if the impurity mass is large enough. 

In order for the anyonic molecule to behave as a rigid object and avoid its internal excitation and its dissociation, one must impose that the cyclotron frequency $\Omega_{\rm cycl}$ is smaller than its lowest excitation mode at $\Delta \omega_{-1}$. This imposes a similar condition
\begin{equation}
    \frac{\mathcal{Q}}{q}\frac{m}{\mathcal{M}}\frac{\omega_{\rm cycl}}{\Delta\omega_{-1}}\ll 1
\end{equation}
that, again, is well satisfied in the heavy impurity limit.
But it is also important to note that forming the bound impurity--quasi-hole state may be itself a non-trivial task since quasi-holes are associated to a global rotation of the FQH fluid. In~\cite{Macaluso_2018,macaluso2020charge}, it was shown that the quasi-hole state naturally forms when the atomic fluid is cooled to its ground state in the presence of the impurity provided that the atoms are able to exchange angular momentum with the external world. Alternatively, a quasi-hole can be created by inserting a localized flux through the cloud, and then introducing the impurity particle at its location~\cite{Grusdt_grow2014,dutta2018coherent}. 
Finally, a speculative strategy yet to be fully explored may consist of inserting the impurity into the FQH fluid through its edge: provided the impurity's motion is slow enough, one can reasonably expect that it will be energetically favourable for the impurity to capture a quasi-hole from the edge and bring it along into the bulk of the FQH cloud.

Besides these technical difficulties, we anticipate that our proposed experiment for the measurement of the fractional charge will have great conceptual advantages over the shot noise measurements of electronic currents that were first used to detect charge fractionalization~\cite{DePicciotto_1997}. These experiments involve in fact complex mechanisms for charge transport and charge injection/extraction into/from the edge of the electron gas. On the other hand, we foresee that our proposed experiment has the potential to provide a direct and unambiguous characterization of the fractional charge of the quasi-hole excitations in the bulk of a fractional quantum Hall fluid.


%
\subsection{Effective Hamiltonian for two anyonic molecules}
\label{sec:TwoImpurities}

After completing the calculation of the single-particle parameters $\mathcal{M}$ and $\mathcal{Q}$, we can now move on to the many-particle case. The two molecule case is already of particular interest as it allows to obtain information about the fractional statistics of the anyonic molecules. In the following we will focus on this case and we will leave the complexities of the three- and more-particle cases~\cite{Wu_1984_2} to future investigations.

As already stated, we assume that the two impurities are located in the bulk of the  FQH cloud, far apart from the edges and they are well separated by a distance much larger than the magnetic length.
The effective molecule Hamiltonian is now given by
\begin{equation}
\begin{split}
    H_{\text{eff}} & 
    =\sum_{j=1}^2\frac{[-i\bold{\nabla}_{\bold{R}_{j}}-Q\bold{A}(\bold{R}_{j})+\bold{\mathcal{A}}_{j}(
    \mathbf{R}
    )]^2}{2\mathcal{M}}\\
    & +\Phi(\bold{R}) + \epsilon_{\rm BO}^{(0)}(\bold{R})+ V_{\text{ii}}(\bold{R})
    \; ,
\end{split}
\label{eq:Ham_two_body}
\end{equation}
where $\mathbf{R}$ is again a shorthand for the whole set of impurity positions, $\{ \mathbf{R}_k\}$. The Berry connection experienced by the $j$-th particle now contains two terms,
\begin{equation}
\begin{split}
    \bold{\mathcal{A}}_{j}(
    \mathbf{R}
    ) & =\bold{\mathcal{A}}_{\text{q}}(\bold{R}_{j})+\bold{\mathcal{A}}_{{\rm stat},j}(
    \mathbf{R})\\
    & =\frac{\mathcal{B}_{\text{q}}}{2} \bold{u}_{\text{z}}\times \bold{R}_{j}
    +(-1)^j\frac{\nu}{R^2_{\text{rel}}}\bold{u}_{\text{z}}\times \bold{R}_{\text{rel}}
    \; .
    \label{eq:Berry_two_terms}
\end{split}
\end{equation}
The former term $\mathcal{A}_{\rm q}$ is of single-particle nature and only depends on the position of the specific particle. Its Berry curvature $\mathcal{B}_{\text{q}}=\nu/\ell_{\text{B}}^2$ accounts for the synthetic magnetic field felt by each quasi-hole via their fractional charge $-\nu q$, as discussed in the previous section. The latter term $\mathcal{A}_{\rm stat}$ has a two-body nature and depends on the relative position of the two impurities,  $\bold{R_{\text{rel}}}=(X_{\text{rel}},Y_{\text{rel}})=\bold{R}_1-\bold{R}_2$: each impurity experiences the vector potential corresponding to $\nu$ quanta of magnetic flux spatially localized on the other impurity. Since $\bold{\nabla}\cross\bold{\mathcal{A}}_{\text{stat},j}=0$, there is no Berry curvature involved in the interaction between spatially separated impurities and the effect can be viewed as an Aharonov-Bohm-like interaction~\cite{Aharonov_1959}. 

Since the impurities are assumed to be located in the bulk of the FQH cloud and to be spatially separated to avoid any overlap, the Born-Oppenheimer energy $\epsilon_{\rm BO}^{(0)}$ does not depend on the positions and can be neglected~\footnote{In the plasma analogy~\cite{Tong_2016}, this is easily understood in terms of the screening of the impurities by the charges which leads to a free energy independent of the impurity’s positions.}. The two-body generalization of the scalar potential $\Phi$ in \eqref{eq:scalarPhi} involves derivatives of the Laughlin wavefunction with respect to the impurity positions~\footnote{The analytic form \eqref{eq:scalarPhi} of the scalar potential only involves the projection of the derivative on orthogonal excited states. As required by gauge invariance, the contribution along the ground state cancels with the vector potential term.}. As it was discussed around \eqref{eq:phi1}, such derivatives only involve localized excitations in the atomic fluid around the quasi-hole. On this basis, for sufficiently separated impurities, we can safely approximate the two-body scalar potential with a relative-coordinate-independent energy shift that can be safely neglected in what follows.

Grouping the single-particle Berry connection due to the effective charge of each quasi-hole with the synthetic magnetic field directly felt by each impurity as done in the previous section, we can write the Hamiltonian in the compact form
\begin{equation}
\label{eq:EffHam2imp}
\begin{split}
    H_{\text{eff}} & 
    =\sum_{j=1}^2\frac{[-i\bold{\nabla}_{\bold{R}_{j}}-\mathcal{Q}\bold{A}(\bold{R}_{j})+\bold{\mathcal{A}}_{\text{stat},j}(\bold{R})]^2}{2\mathcal{M}}\\
    & +V_{\text{ii}}(\bold{R})\; ,
\end{split}
\end{equation}
in terms of the effective charge $\mathcal{Q}=Q-\nu q$ of each molecule. According to this Hamiltonian, the molecules interact via the interaction potential $V_{\text{ii}}$ between the bare impurities and via the Aharonov-Bohm interaction encoded by the  two-body vector potential $\bold{\mathcal{A}}_{\text{stat},j}$ that depends 
on the relative position $\bold{R}_{1}-\bold{R}_{2}$ between the two molecules.

Given the translational invariance of the configuration, we can separate the center of mass and the relative motion of the two molecules. 
Assuming a central impurity-impurity interaction $V_{\text{ii}}(\bold{R}_{i}-\bold{R}_{j})=V_{\text{ii}}(R_{\text{rel}})$, we define the reduced and the total mass as usual as
\begin{align}
    \mathcal{M}_{\text{rel}}=\mathcal{M}/2 \; , \; \;
    \mathcal{M}_{\text{CM}}=2\mathcal{M} \; ,
\end{align}
the relative and center of mass position
\begin{eqnarray}
\bold{R}_{\text{rel}}&=&\bold{R}_1-\bold{R}_2\; , \\
\bold{R}_{\text{CM}}&=&\frac{\bold{R}_1+\bold{R}_2}{2}\; ,
\end{eqnarray}
the corresponding momenta
\begin{eqnarray}
\bold{P}_{\text{rel}}&=&\frac{\bold{P}_1-\bold{P}_2}{2}\; , \\ \bold{P}_{\text{CM}}&=&\bold{P}_1+\bold{P}_2\; ,
\end{eqnarray}%
and vector potentials
\begin{eqnarray}
\bold{A}_{\text{rel}}(\bold{R}_{\text{rel}})&=&\frac{\mathcal{Q}}{2} \bold{A}(\bold{R}_{\text{rel}}) +
    \frac{\bold{\mathcal{A}}_{\text{stat},1}(\bold{R}_1)-\bold{\mathcal{A}}_{\text{stat,2}}(\bold{R}_2)}{2} = \nonumber \\
    &=&\frac{\mathcal{Q}}{2} \bold{A}(\bold{R}_{\text{rel}}) + \frac{\nu}{R_{\text{rel}}^2}(-Y_{\text{rel}},X_{\text{rel}})\; , \\ 
    \bold{A}_{\text{CM}}(\bold{R}_{\text{CM}})
    &=&  
    2\mathcal{Q}\,\bold{A}(\bold{R}_{\text{CM}})\; ,
\end{eqnarray}
to be included in the center of mass and relative Hamiltonians
\begin{align}
    \label{eq:HCM}
    &H_{\text{CM}}=\frac{\left[\bold{P}_{\text{CM}}- \bold{A}_{\text{CM}}(\bold{R}_{\text{CM}})\right]^2}{2\mathcal{M}_{\text{CM}}} \; ,\\
    \label{eq:Hrel}
    &H_{\text{rel}}=\frac{\left[\bold{P}_{\text{rel}}+\bold{A}_{\text{rel}}(\bold{R}_{\text{rel}})\right]^2}{2\mathcal{M}_{\text{rel}}}+V_{\text{ii}}(R_{\text{rel}})\; .
\end{align}
%
The center of mass Hamiltonian \eqref{eq:HCM} describes a free particle motion of total mass $2\mathcal{M}$ and charge $2\mathcal{Q}$. On the other hand, the relative Hamiltonian \eqref{eq:Hrel} contains the uniform magnetic field experienced by the reduced charge $\mathcal{Q}/2$ plus 
a non-trivial vector potential corresponding to $\nu$ quanta of magnetic flux localized at $R_{\text{rel}}=0$.

As it was discussed in the seminal works~\cite{Leinaas_1977,Wilczek_1982,Halperin_1984,Arovas_1984,Wu_1984,Wu_1984_2}, the presence of this latter vector potential is the key feature that encodes the fractional statistics of the anyonic molecules. In the following of this work, we will study the effect of this vector potential onto the scattering cross section of two molecules. This is a measurable quantity that can serve as a probe of the  statistical parameter of the molecules.

Depending on the bosonic vs. fermionic nature of the impurities, the effective Hamiltonian \eqref{eq:EffHam2imp} will act on the Hilbert space of symmetric or anti-symmetric wavefunctions under the exchange of the two molecules, that is $\mathbf{R}_1\leftrightarrow \mathbf{R}_2$ (or $\mathbf{R}_{\rm rel}\leftrightarrow -\mathbf{R}_{\rm rel}$).
The combination of the intrinsic statistics of the impurities and the one inherited by the quasi holes can be encoded in the single statistical parameter  $\alpha=\alpha_{\text{i}}+\nu$, where the intrinsic contribution is $\alpha_{\text{i}}=0$ ($\alpha_{\text{i}}=1$) for bosonic (fermionic) impurities. In the next section, we will see how the scattering properties only depend on $\alpha$ and not on $\alpha_{\rm i}$ and $\nu$ separately.

%
\section{Scattering of anyonic molecules and fractional statistics}
\label{sec:TwoImpuritiesAnyPolScatt}
\begin{figure}
    \centering
    \includegraphics[width=0.5\textwidth]{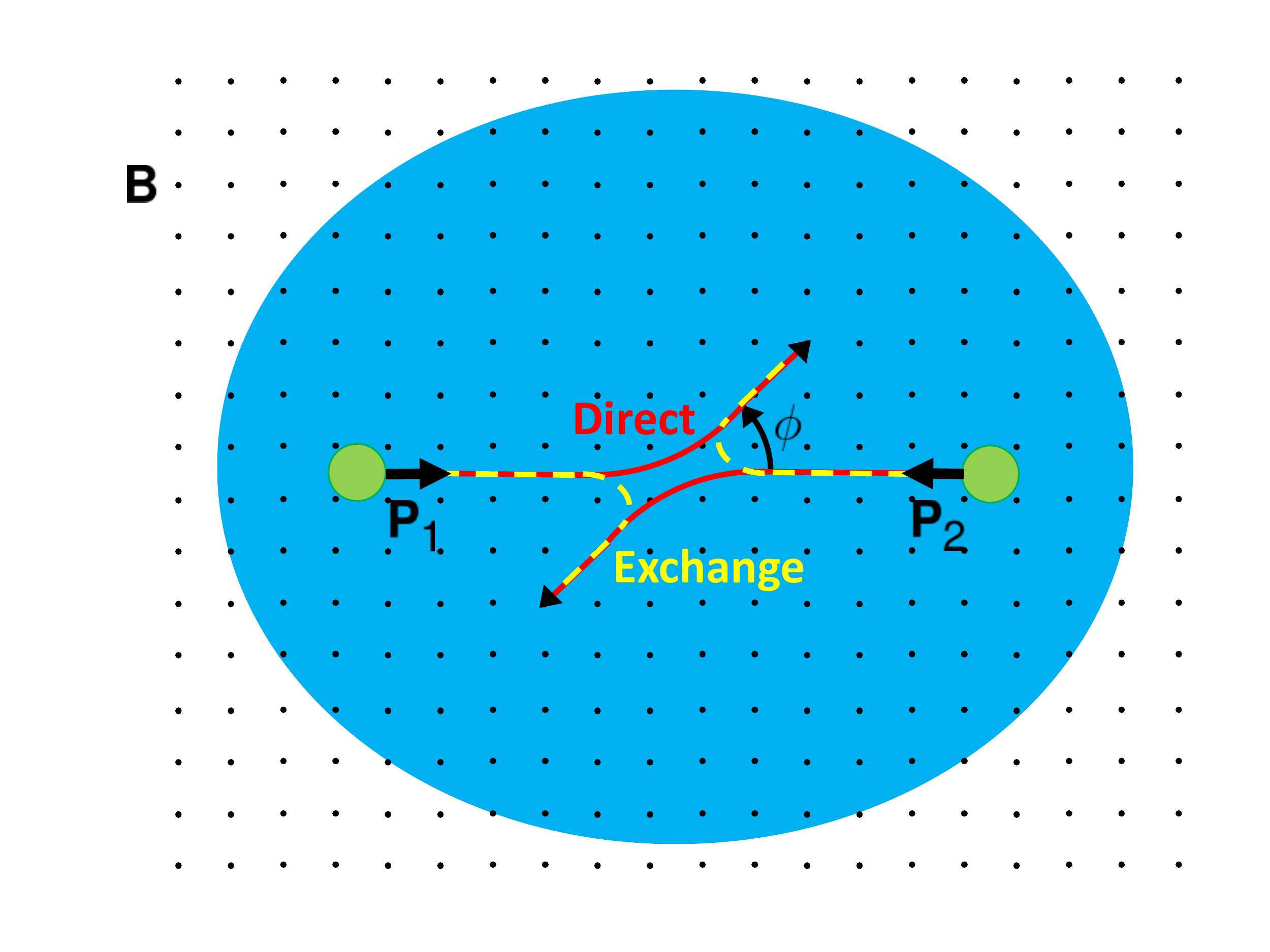}
    \caption{{\em Scattering of two anyonic molecules.} Similarly to Fig.~\ref{fig:ExpSingleImp}, two indistinguishable anyonic molecules (green circles) formed by the binding of the same number of quasi-holes to a pair of identical impurities in the bulk of a FQH fluid (blue region) are considered. The two molecules are given momentum kicks against each other ($\bold{P}_1$ and $\bold{P}_2$, respectively). Due to their indistinguishability, two scattering channels contribute to the differential scattering cross section at an angle $\phi$, see Eq.~\eqref{eq:DifScattCrossSecIdentical}: the two channels are labelled as \textit{direct} (red, solid trajectories) and \textit{exchange} (yellow, dashed ones) and involve a relative phase determined by the anyonic statistics. As one can guess from textbook two-slit interference, information about the statistics can be extracted from the global position of the interference fringe pattern. This is illustrated in the next figures.}
    \label{fig:ExpScattering}
\end{figure}

In the previous Section, we have summarized the conceptual framework to study the quantum mechanical motion and the interactions of anyonic molecules. Based on this complete and flexible framework, we can now attack the core subject of this work, namely the observable consequences of the fractional statistics. As a simplest and most exciting example, we consider the differential cross section for the scattering of two anyonic molecules and, in particular, we will highlight a simple relation between the angular position of its maxima and minima and the fractional statistics.

To simplify our discussion, from now on we assume that the process underlying the synthetic magnetic field is designed in a way to have a vanishing effective charge $\mathcal{Q}=0$ of the molecule. This condition is beneficial to have rectilinear trajectories in the asymptotic states of the scattering molecules.
The only vector potential remaining in Eq.~\eqref{eq:EffHam2imp} will then be the Aharonov-Bohm interaction $\bold{\mathcal{A}}_{\text{rel}}$, which simplifies enormously the study of the scattering process.

A scheme of the proposed experimental strategy can be found in Fig.~\ref{fig:ExpScattering}. If one prepares a pair of identical molecules inside the bulk of the FQH droplet, each one composed of the same kind of impurity and a bound quasi-hole excitation, and then makes them to collide, e.g. by pushing them against each other via a suitable external potential, the angular dependence of the differential scattering cross section will show a pattern of maxima and minima whose angular position can be directly related to the fractional statistics of the molecules, as we will show in Sec.~\ref{sec:Numerics}.

%
\subsection{General scattering theory}

In order to study the two-molecule scattering we focus on the relative Hamiltonian \eqref{eq:Hrel} in 2D cylindrical coordinates and we consider the time-independent Schr\"{o}dinger equation
\begin{equation}
\label{eq:SErel}
\begin{split}
&\Bigg[\frac{\partial^2}{\partial r^2}+\frac{1}{r}\frac{\partial}{\partial r}+\frac{1}{r^2}\left(\frac{\partial}{\partial\phi}-i\nu\right)^2\!\!
-2\mu V_{\text{ii}}(r)+k^2\Bigg]\psi (r,\phi)\\ & =0 \;,
\end{split}
\end{equation}
where $k^2=2\mu E$ is related to the energy $E$ of the scattering process. For the sake of notational simplicity we use the shorthand $r,\mu$ in place of $R_{\text{rel}},\mathcal{M}_{\text{rel}}$. Eq.~\eqref{eq:SErel} represents the scattering of a particle of mass $\mu$ by a flux tube of radius $r_0\rightarrow 0$ giving a vector potential $\bold{A}_{\text{rel}}=\nu\bold{u}_\phi/r$ that incorporates the fractional statistics ($\bold{u}_\phi$ is a unit vector in the $\phi$ coordinate).
For a short-range potential (i.e. such that $rV_{\text{ii}}(r)\rightarrow 0$ when $r\rightarrow\infty$) the solution far from the origin can be written as the sum of an incoming plane wave~\footnote{It has been argued that in order to have a constant incoming current density one should have an incoming plane wave of the form $e^{i\bold{k}\cdot\bold{r}}e^{i\nu\phi}$ with a non-trivial angular dependence of the phase. Nevertheless, the only consequence of the absence of the phase factor  is the appearance in $f(k,\phi)$ of a $\delta$-function addend in the forward direction, namely around $\phi=0$~\cite{Hagen_1990}. Moreover, as $r\rightarrow\infty$ the incorrect incoming current density calculated with the plane wave in Eq.~\eqref{eq:solutionSEinout} tends to the correct, constant value that one would obtain if the factor $e^{i\nu\phi}$ was correctly included~\cite{Caenepeel_1994}. On this basis, in what follows we will for simplicity ignore both the correction to the incoming plane wave and the forward $\delta$-function whose absence leads to.} 
and an outgoing cylindrical wave~\cite{Lapidus_1982,Adhikari_1986}
\begin{align}
\label{eq:solutionSEinout}
\psi(\bold{r})=e^{i\bold{k}\cdot\bold{r}}+f(k,\phi)\frac{e^{ikr}}{\sqrt{r}} \;,
\end{align}
where $f(k,\phi)$ is the scattering amplitude.

We will solve Eq.~\eqref{eq:SErel} using the method of partial waves.
Given the cylindrical symmetry of the problem, we can look for factorized solutions $\psi(r,\phi)=e^{im\phi}\,u_{m,\nu}(r)/\sqrt{r}$ of angular momentum $m$ with a radial function satisfying
\begin{equation}
\label{eq:SEu}
\begin{split}
&\frac{d^2u_{m,\nu}}{dr^2}+\frac{u_{m,\nu}(r)}{4r^2}-\frac{\left(m-\nu\right)^2}{r^2}u_{m,\nu}(r)\\ &-2\mu V_{\text{ii}}(r)u_{m,\nu}(r)+k^2 u_{m,\nu}(r) =0\;.
\end{split}
\end{equation}
In contrast to usual scattering problems, for any non-integer value of $\nu$ the centrifugal barrier is present here for all values of $m$. This guarantees that the wave function vanishes for $r=0$ when the two particles overlap~\cite{Wu_1984_2}.

The general solution of Eq.~\eqref{eq:SEu} in the free case $V_{\text{ii}}=0$ with $\nu=0$ has the form
\begin{align}
u_{m}(r)\propto { \sqrt{r}}\,J_{m}(kr) \; ,
\end{align}
in terms of the cylindrical Bessel function $J_{m}(kr)$. In the $r\rightarrow\infty$ limit the expression above tends to  
\begin{align}
\label{eq:AsymptoticFreeSolution}
u_{m}(r)=\sqrt{\frac{2}{\pi k}}\cos{\left(kr-m\frac{\pi}{2}-\frac{\pi}{4}\right)} \; .
\end{align}
For any short-range potential $V_{\text{ii}}$ the solution of Eq.~\eqref{eq:SEu} in the $r\rightarrow\infty$ limit can be written w.r.t. the asymptotic form of the free solution~\eqref{eq:AsymptoticFreeSolution} as
\begin{align}
\label{eq:u_asymp}
u_{m,\nu}(r)=A_{m,\nu}(k)\cos\left[kr+\delta_{m,\nu}(k)-m\frac{\pi}{2}-\frac{\pi}{4}\right] \;,
\end{align}
where $\delta_{m,\nu}$ are the phase shifts.

As usual, the scattering amplitude in Eq.~\eqref{eq:solutionSEinout} can be related to the phase shifts $\delta_{m,\nu}(k)$ of this asymptotic expansion: using the fact that the cylindrical harmonics are a complete basis and replacing all cylindrical Bessel functions with their asymptotic form at $r\rightarrow\infty$, we can write the wavefunction in Eq.~\eqref{eq:solutionSEinout} as:
\begin{equation}
\label{eq:WfLargeR}
\begin{split}
&\psi(\mathbf{r})=\left[\sum_{m=-\infty}^{+\infty}i^m \sqrt{\frac{2}{\pi kr}}\cos{\left(kr-m\frac{\pi}{2}-\frac{\pi}{4}\right)}e^{im\phi}\right]
\\&+\left[\sum_{m=-\infty}^{\infty}a_{m,\nu}(k)e^{im\phi}\right]\frac{e^{ikr}}{\sqrt{r}}
\\&=\sum_{m=-\infty}^{+\infty}A_{m,\nu}(k)\cos\left[kr+\delta_{m,\nu}(k)-m\frac{\pi}{2}-\frac{\pi}{4}\right]\frac{e^{im\phi}}{\sqrt{r}} \;,
\end{split}
\end{equation}
with 
\begin{equation}
    A_{m,\nu}(k)=\sqrt{\frac{2}{\pi k}}i^m e^{i\delta_{m,\nu}(k)}\; ,
\end{equation}
from which it is easy to obtain an expression of the scattering amplitude in terms of the phase shifts
\begin{align}
\label{eq:f_sum_a}
    f(k,\phi)=\sum_{m=-\infty}^{\infty}a_{m,\nu}(k)e^{im\phi}\; ,
\end{align}
with
\begin{equation}
a_{m,\nu}(k)=\sqrt{\frac{2i}{\pi k}}e^{i\delta_{m,\nu}(k)}\sin{\delta_{m,\nu}(k)}\;. \label{eq:a_func_delta}
\end{equation}
%

%
\subsection{A general result for short-range potentials}

For a non-vanishing and non-integer $\nu$ the free solution of Eq.~\eqref{eq:SEu} changes to
\begin{align}
\label{eq:uFreeNoA}
u_{m,\nu}(r)\propto { \sqrt{r}}\,J_{|m-\nu|}(kr) \;,
\end{align}
which approaches
\begin{align}
u_{m,\nu}(r)=C\sqrt{\frac{2}{\pi k}}\cos{\left[kr-|m-\nu|\frac{\pi}{2}-\frac{\pi}{4}\right]}
\label{eq:u_nu_noV}
\end{align}
in the $r\rightarrow\infty$ limit.

For any short-range potential $V_{\text{ii}}$, the total phase shift in the cosine $\cos(kr+\Delta)$ in the asymptotic limit $r\rightarrow\infty$ can be referred to the fully free case with $V_{\text{ii}}=0$ and $\nu=0$ [as done for $\delta_{m,\nu}(k)$ in Eq.~\eqref{eq:u_asymp}] or to the non-interacting case $V_{\text{ii}}=0$ with $\nu\neq 0$ considered in \eqref{eq:u_nu_noV}. These two choices give
\begin{align}
&\Delta=\delta_{m,\nu}(k)-m\frac{\pi}{2}-\frac{\pi}{4} \; , \\
&\Delta=\Delta^V_{{m,\nu}}-|m-\nu|\frac{\pi}{2}-\frac{\pi}{4} \; ,
\end{align}
respectively, where $\Delta^V_{m,\nu}$ is the phase shift exclusively due to the intermolecular potential $V_{\text{ii}}$. Combining these equations we obtain
\begin{align}
\label{eq:DeltaDecomp}
\delta_{m,\nu}(k)=m\frac{\pi}{2}-|m-\nu|\frac{\pi}{2}+\Delta^V_{m,\nu}(k) \; ,
\end{align}
where the total phase shift $\delta_{m,\nu}(k)$ is decomposed as the sum of the phase shift due to the topological flux attached to the impurities plus the one $\Delta^V_{m,\nu}(k)$ due to the interaction potential. In the non-interacting $V_{\text{ii}}=0$ case, this yields the same result as calculated in the original work by Aharonov and Bohm~\cite{Aharonov_1959}.

In the general case, assuming $0< \nu <1$, we can combine Eqs.~\eqref{eq:a_func_delta} and~\eqref{eq:DeltaDecomp} and decompose the scattering amplitude as
\begin{multline}
\label{eq:f_final}
f(k,\phi)=f_{\text{AB}}(k,\phi)+f_{\text{V}}(k,\phi)= \\ 
=\frac{1}{\sqrt{2\pi ik}}\Bigg\{\Bigg[
\sum_{m=1}^{\infty} e^{im\phi}\left(e^{i\pi\nu}-1\right)\\+\sum_{m=-\infty}^{0} e^{im\phi}\left(e^{-i\pi\nu}-1\right)\Bigg] + \\
+\Bigg[e^{i\pi\nu}\sum_{m=1}^{\infty}e^{im\phi}\left(e^{2i\Delta_{m,\nu}^{\text{V}}}-1\right)+\\
+e^{-i\pi\nu}\sum_{m=-\infty}^{0}e^{im\phi}\left(e^{2i\Delta_{m,\nu}^{\text{V}}}-1\right)\Bigg]\Bigg\} \;.
\end{multline}
The terms on the first two lines are geometric series that can be analytically summed up to $m=\infty$. They give the Aharonov-Bohm contribution to the scattering amplitude~\cite{Aharonov_1959}
\begin{equation}
\label{eq:ABf}
f_{\text{AB}}(k,\phi)=\sum_{m=-\infty}^{+\infty}a^{\text{(AB)}}_{m,\nu}(k)e^{im\phi}=-\frac{\sin{(\pi\nu)}}{\sqrt{2\pi ik}}\frac{e^{i\phi/2}}{\sin{\frac{\phi}{2}}} \;,
\end{equation}
and carry all information on the particle statistics.
The terms on the third and fourth line summarize instead the contribution $f_{\text{V}}(k,\phi)$ of the interaction potential $V_{\text{ii}}$ to the scattering amplitude. These terms must be evaluated by numerically summing the series. 

This decomposition is of crucial technical importance as it enables to isolate the Aharonov-Bohm contribution $f_{\text{AB}}$ that can be analytically computed, and restrict the numerical calculation to the potential contribution $f_{\text{V}}$ only, for which convergence on the high angular-momentum side is straightforward. 
This is however much more than just a mathematical trick, since it tells us about the different physical nature of the two contributions to the scattering amplitude. The statistical part of the scattering amplitude $f_{\text{AB}}$ originates from a vector potential $\bold{\mathcal{A}}_{\text{rel}}$ that extends to infinity. As a result, it affects all angular momentum components. Its divergent behaviour for $\phi\to 0$ can be physically related to the step-like jump of the geometric phase that is accumulated when passing in the close vicinity of $r=0$ on opposite sides. On the other hand, for a short-range interaction potential, the particles only see each other up to a certain distance, and therefore one only needs to sum up to a finite number of partial waves to achieve convergence in $f_{\text{V}}$. 

%
\subsection{Distinguishable and indistinguishable impurities}

For distinguishable impurities, the differential scattering cross section is calculated directly from the scattering amplitude as
\begin{equation}
\label{eq:ds_disting}
    \frac{d\sigma_{\text{D}}}{d\phi}=|f(k,\phi)|^2\; .
\end{equation}
Nevertheless, the scattering process is most interesting when the impurities are indistinguishable particles. In this case, the differential scattering cross section involves a sum over exchange processes according to
\begin{align}
\label{eq:DifScattCrossSecIdentical}
    \frac{d\sigma_{\text{B},\text{F}}}{d\phi}=|f(k,\phi)\pm f(k,\phi+\pi)|^2 \; ,
\end{align}
and may thus allow for interesting interference features in the angular dependence.
As usual, the $\pm$ signs here correspond to bosonic and fermionic impurities, respectively. Indistinguishability guarantees that the cross section has the same value for $\phi$ and $\phi+\pi$.
%

Repeating the same calculation leading to \eqref{eq:f_final} in the $1<\nu<2$ case and noting that $\Delta_{m,\nu}^{\text{V}}$ is a function of $m-\nu$ only, one can show that 
\begin{equation}
   f_{1+{\nu}}(k,\phi)=-e^{i\phi}\,f_{\nu}(k,\phi)
\end{equation}
holds for any $0<\nu<1$.
It is then immediate to deduce that the scattering cross sections in the bosonic and fermionic cases are related by~\footnote{Extending our theory to the  $\nu=1$ integer quantum Hall (IQH) regime, the relation \eqref{eq:BoseFermi} has the direct physical interpretation that a boson immersed in the IQH transmutes into a fermion upon binding a quasi-hole, which in this case corresponds to one missing particle in the underlying IQH fluid.}
\begin{equation}
\frac{d\sigma_{\text{B},1+{\nu}}}{d\phi}(\phi)=
\frac{d\sigma_{\text{F},\nu}}{d\phi}(\phi)\;.
\label{eq:BoseFermi}
\end{equation}
Noting the scattering cross sections are periodic of period 2 in $\nu$, one can thus summarize the statistics into a single statistical parameter $\alpha$, defined as $\alpha=\alpha_{\rm i}+\nu$ with $\alpha_{\rm i}=0$ ($\alpha_{\rm i}=1$) for bosonic (fermionic) impurities, which fully determines the scattering properties as a general scattering cross section $d\sigma_\alpha/d\phi$.

A similar reasoning leads to the interesting relation
\begin{equation}
    f_{1-\nu}(k,\phi)=-e^{i\phi} f_\nu(k,-\phi)\;,
\end{equation}
from which one extracts the symmetry relation
\begin{equation}
\frac{d\sigma_{\text{B},\nu}}{d\phi}(\pi-\phi)=
\frac{d\sigma_{\text{F},1-\nu}}{d\phi}(\phi)\,\;,
\label{eq:Bose-Fermi1}
\end{equation}
that translates into the compact form
\begin{equation}
\frac{d\sigma_{\alpha}}{d\phi}(\pi-\phi)=
\frac{d\sigma_{2-\alpha}}{d\phi}(\phi)\,\;.
\label{eq:Bose-Fermi2}
\end{equation}

\subsection{Numerical results for the differential scattering cross section}
\label{sec:Numerics}
\begin{figure}
    \centering
    \includegraphics[width=0.5\textwidth]{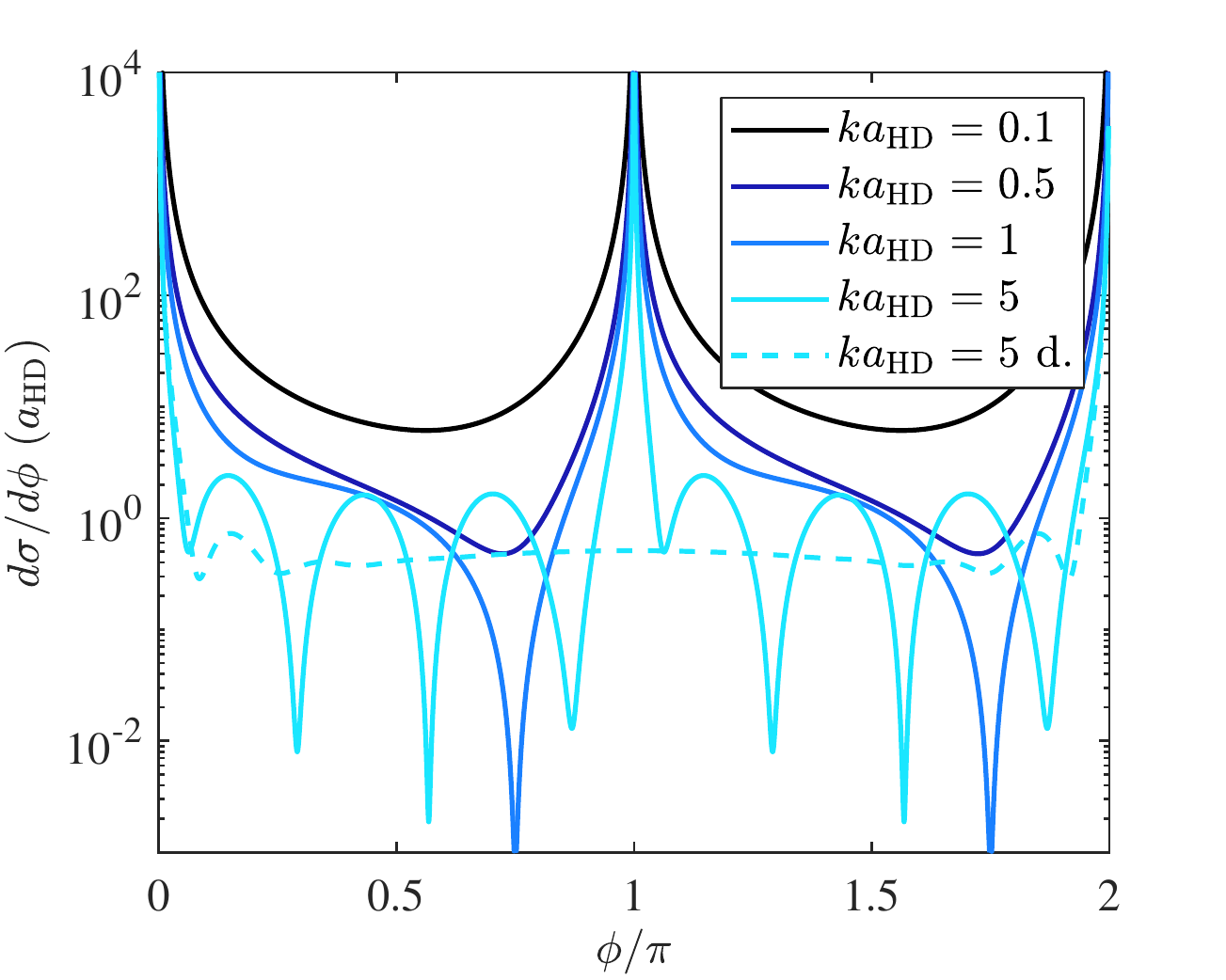}
    \caption{Differential scattering cross section $d\sigma/d\phi$ in units of the hard-core radius $a_{\rm HD}$ as function of the scattering angle $\phi$ for a hard-disk interaction potential  $V_{\text{ii}}$ between bosonic impurities immersed in a $\nu=0.5$ FQH fluid. The different solid curves correspond to indistinguishable impurities with different values of the relative incident momentum $ka_{\rm HD}$. The larger $ka_{\rm HD}$, the more visible is the fringe pattern resulting from the interference of direct and exchange scattering channels. For comparison, the dashed curve is for distinguishable impurities at $ka_{\rm HD}=5$: in this case, no interference fringe is visible.}
    \label{fig:HD_alpha05}
\end{figure}
\begin{figure}
    \centering
    \includegraphics[width=0.5\textwidth]{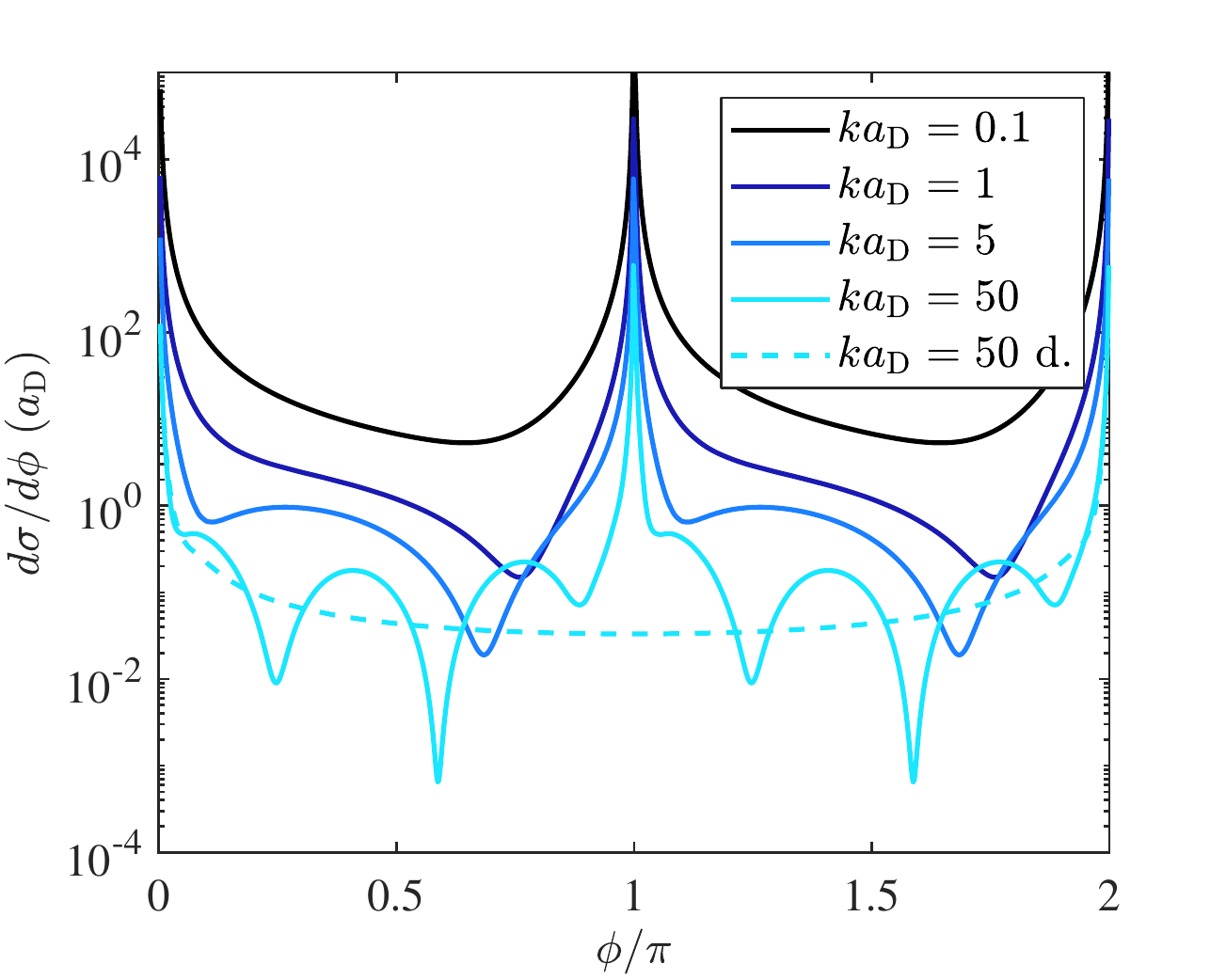}
    \caption{Differential scattering cross section $d\sigma/d\phi$ in units of the dipolar length $a_{\rm D}$ as function of the scattering angle $\phi$ for a dipole interaction potential  $V_{\text{ii}}$ between bosonic impurities immersed in a $\nu=0.5$ FQH fluid. The different solid curves correspond to indistinguishable impurities with different values of the relative incident momentum $ka_{\rm D}$. The larger $ka_{\rm D}$, the more visible is the fringe pattern resulting from the interference of direct and exchange scattering channels. For comparison, the dashed curve is for distinguishable impurities at $ka_{\rm D}=50$: in this case, no interference fringe is visible.}
    \label{fig:Dipole_alpha05}
\end{figure}
\begin{figure}
    \centering
    \includegraphics[width=0.5\textwidth]{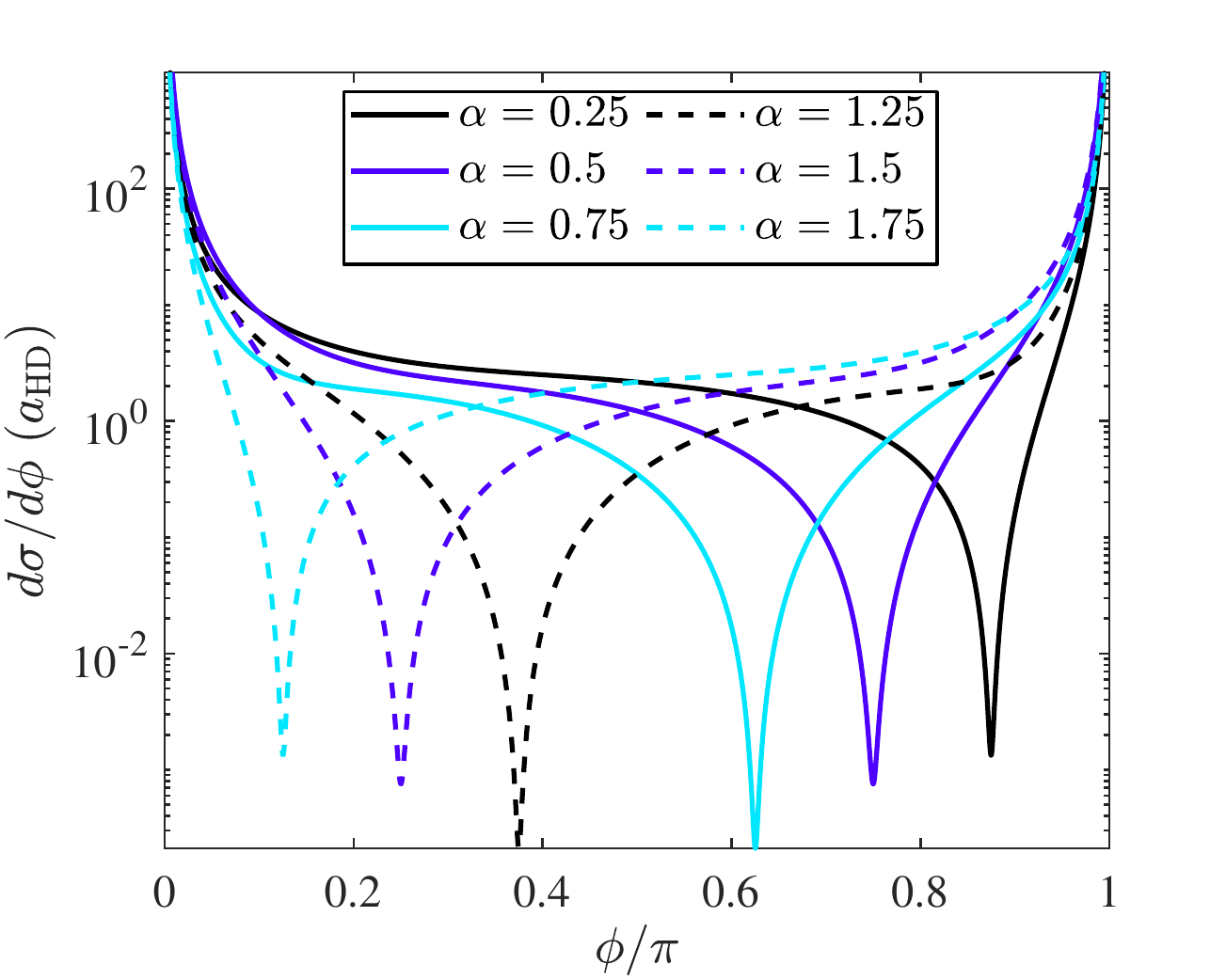}
    \caption{Differential scattering cross section $d\sigma/d\phi$ in units of the hard-core radius $a_{\rm HD}$ as function of the scattering angle $\phi$ for a hard-disk interaction potential $V_{\text{ii}}$ between indistinguishable impurities, a relative incident momentum $ka_{\rm HD}=1$ and different values of the statistical parameter $\alpha$, defined as $\alpha=\nu$ for bosons and $\alpha=1+\nu$ for fermions. Solid  (dashed) lines correspond to bosonic (fermionic) impurites inside a FQH bath of filling fraction $0<\nu<1$.}
    \label{fig:BosonsFermions}
\end{figure}
\begin{figure*}
    \centering
    \includegraphics[width=\textwidth]{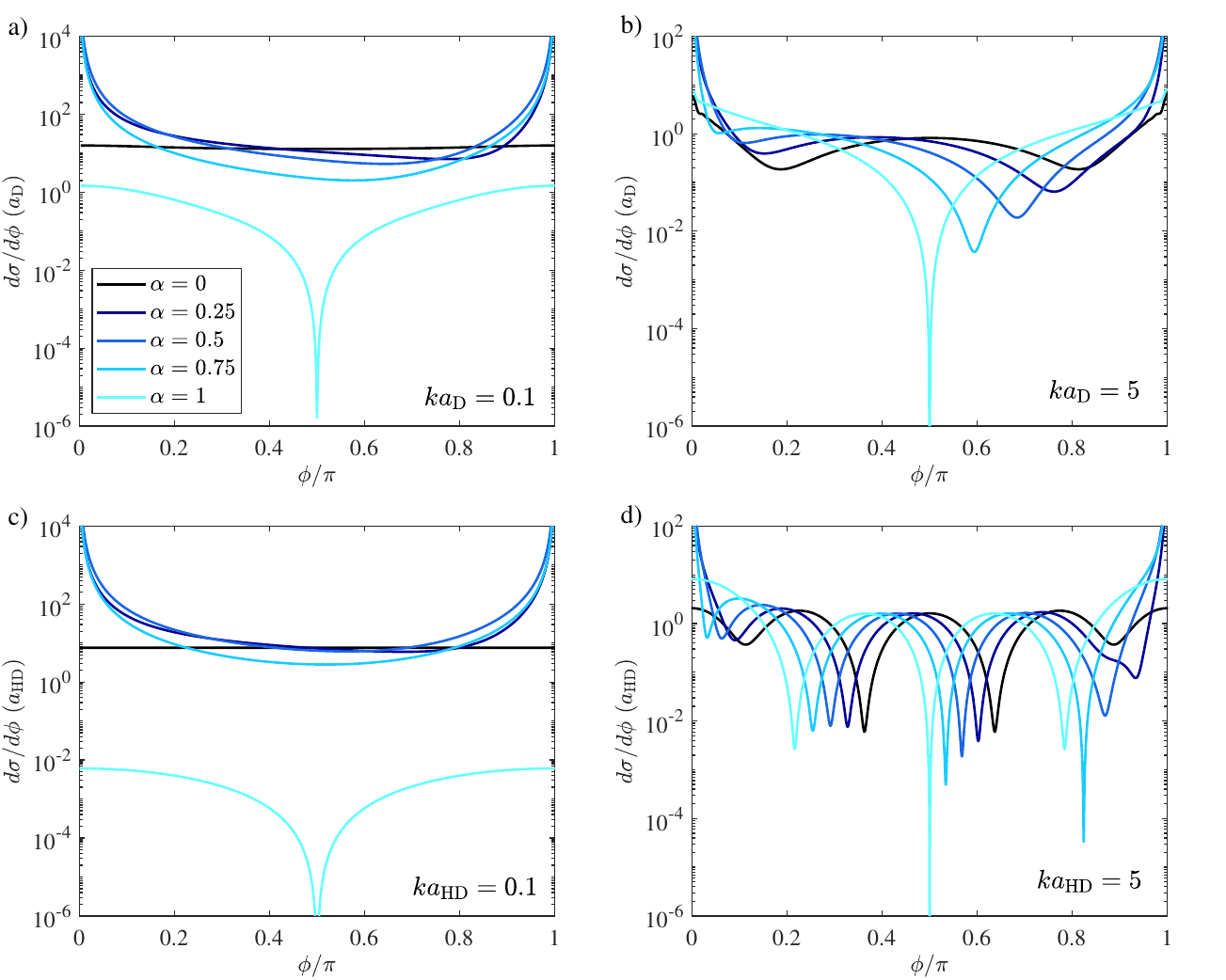}
    \caption{Differential scattering cross section $d\sigma/d\phi$ as function of the scattering angle $\phi$ for several values of the statistical parameter $\nu$. The different panels refer to different values of the relative incident momentum $k$ and different forms of the interaction potential $V_{\text{ii}}$ between the impurities. In particular: \textbf{a)} Dipolar interactions and $ka_{\rm D}=0.1$. \textbf{b)} Dipolar interactions and  $ka_{\rm D}=5$ \textbf{c)} Hard-disk interactions and $ka_{\rm HD}=0.1$ \textbf{d)} Hard-disk interactions and $ka_{\rm HD}=5$. The different curves refer to different values of the statistical parameter as indicated in the legend in a): $\alpha=0, 0.25, 0.5, 0.75$ correspond to bosonic impurities at growing values of $\nu$, while $\alpha=1$ is the fermionic case with $\nu=0$. To avoid overcrowding the figure, no other curve for fermionic impurities is displayed. As it was shown in Fig.~\ref{fig:BosonsFermions}, such curves are immediately obtained from the bosonic ones via the symmetry relations in Eqs.~\eqref{eq:Bose-Fermi1}-\eqref{eq:Bose-Fermi2}.}
    \label{fig:HD_Dipolar_kaFixed}
\end{figure*}
\begin{figure}
    \centering
    \includegraphics[width=0.5\textwidth]{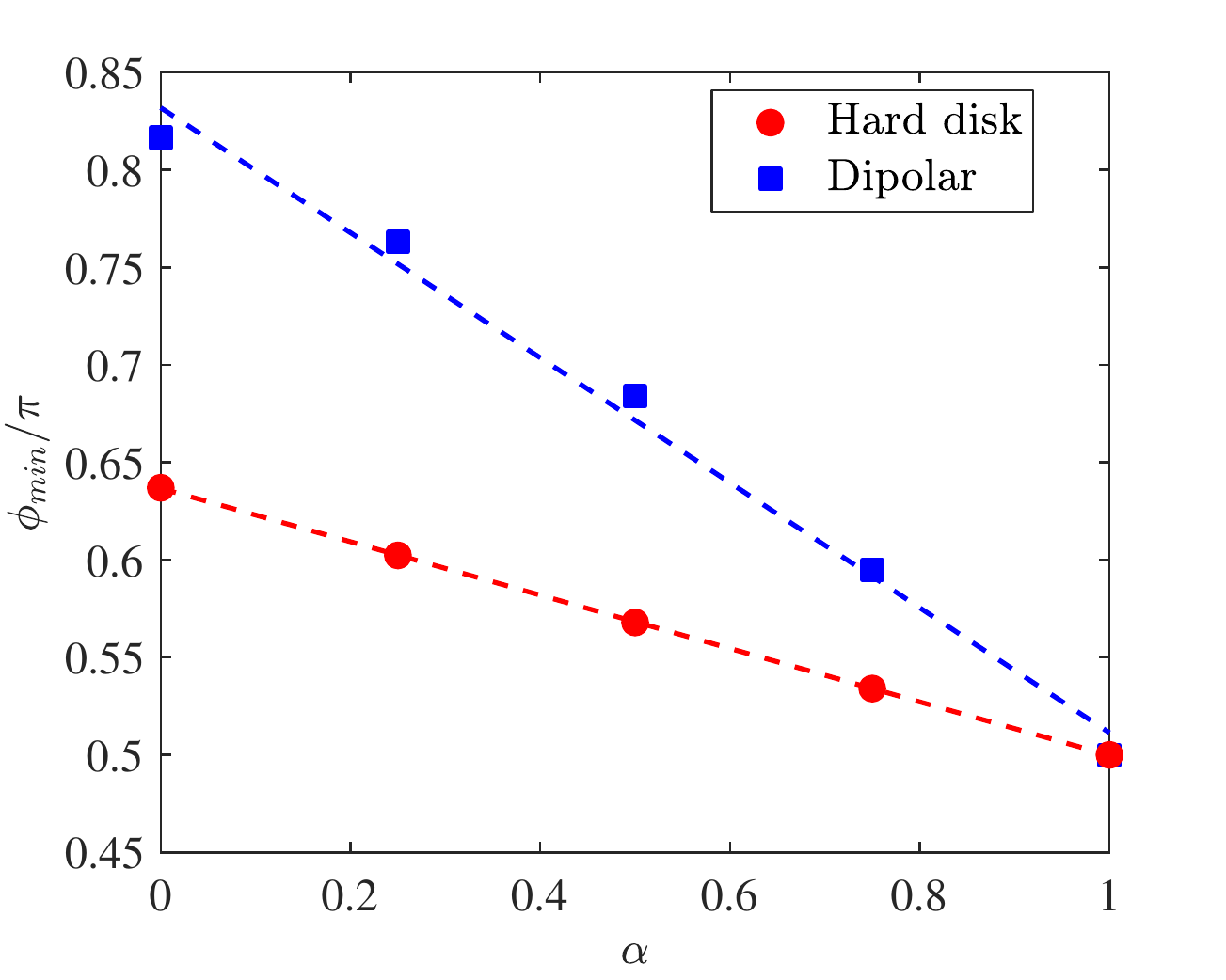}
    \caption{Angular position $\phi_{\text{min}}$ of the minimum of the differential scattering cross section extracted from Fig.~\ref{fig:HD_Dipolar_kaFixed}\textbf{b)},\textbf{d)} for a dipolar and a hard-disk interaction potential, respectively, as a function of the statistical parameter $\alpha=\nu$ of anyonic molecules formed by bosonic impurities bound to a single quasi-hole in a FQH fluid at filling $\nu$. The dashed lines are linear fits. 
    }
    \label{fig:alphaLinear}
\end{figure}
The key feature of the Aharonov-Bohm contribution \eqref{eq:ABf} to the scattering amplitude in the absence of interaction potential $V_{\text{ii}}$ is a divergent behaviour in the forward direction for any non-integer $\nu$. This was traced back by Ref.~\cite{Sommerfield_2000} to the infinite-range nature of the Aharonov-Bohm interaction and is known to pose mathematical difficulties related to the optical theorem. In the following of our work, we are going to focus on the scattering at finite angles $\phi$ where such problems do not arise. 

The situation gets much more interesting when the interaction potential $V_{\text{ii}}$ is included. This introduces a more complex angular dependence of the scattering amplitude $f_V(k,\phi)$ and clear features in the differential scattering cross section. Keeping an eye on possible experimental realizations of this work we choose a dipolar form of the repulsive interaction potential~\cite{Lahaye_2009}, $V_{\text{ii}}=b/r^3$, for which a dipolar length can be defined as $a_{\rm D}=\mu b$. As we shall see better in Sec.~\ref{sec:expt_scatt}, this specific choice of the interaction potential is motivated by the extremely strong dipolar potentials that can be obtained using heteronuclear molecules.
As an additional check, we have also calculated the differential scattering cross section for the case of a hard-wall potential of radius $a_{\rm HD}$,
\begin{equation}
V(r)= \left\{ \begin{array}{lcc}
             \infty &   if  & r \leq a_{\rm HD} \\
             \\ 0 &  if & r>a_{\rm HD}
             \end{array}
   \right.  \; ,
\end{equation}
for which we can benchmark our predictions against the semi-analytical results available in Ref.~\cite{Suzuki_1991}.

In order to calculate the differential cross section $d\sigma/d\phi$ in the different cases, we have to first calculate the phase shift $\delta_{m,\nu}(k)$ by solving the radial Schr\"odinger equation Eq.~\eqref{eq:SEu} in the presence of the interaction potential $V_{\text{ii}}$  and the vector potential $\bold{A}_{\text{rel}}$. This was done numerically, employing Numerov's method~\cite{Numerov_1924}, which gives a global error of order $\mathcal{O}(h^4)$, being $h=r_{i+1}-r_{i}$ the numerical step size in the $r$ coordinate. 
Identifying the interaction contribution $\Delta^V_{m,\nu}$ to the phase shift, one can separate the different terms in Eq.~\eqref{eq:f_final}: the first two lines are analytically computed giving $f_{\text{AB}}$ of Eq.~\eqref{eq:ABf}. The interaction-induced amplitude $f_{\text{V}}$ is evaluated by numerically performing the sum in the last two lines up to large values of $m$ until convergence is reached. The desired differential cross section $d\sigma/d\phi$ is finally obtained by summing the resulting $f_{\text{V}}$ to the analytically computed $f_{\text{AB}}$ and plugging the outcome into Eq.~\eqref{eq:DifScattCrossSecIdentical}.

Figs.~\ref{fig:HD_alpha05} and~\ref{fig:Dipole_alpha05} show the differential scattering cross section as function of the scattering angle $\phi$ for a hard-disk and a dipolar interaction between bosonic impurities, respectively. The filling fraction of the FQH bath is fixed in both cases to $\nu=0.5$, while the different solid curves represent different values of the relative incident momentum $ka_{\rm HD,D}$ for indistinguishable impurities. The qualitative behavior of $d\sigma/d\phi$ is identical for the two potentials: for small momenta ($ka_{\rm HD,D}\ll 1$) the only effect of the fractional statistics beyond the divergences in $\phi=0,\pi$ is the slight breaking of the $\phi\leftrightarrow \pi-\phi$ symmetry (or, equivalently, of the $\phi \leftrightarrow  -\phi$ symmetry), so that the minimum of the curve is displaced to angles larger than $\phi=\pi/2$, as was first found in Ref.~\cite{Suzuki_1991}. 
For larger momenta $ka_{\rm HD,D}\gtrsim 1$, the peaks around $\phi=0,\pi$ persist, and marked oscillations appear in the angular dependence because of interference effects, with a strong suppression of the differential scattering cross section occurring at some particular angles $\phi_{\text{min}}$. 

This oscillating behavior is reached for smaller $ka_{\rm HD,D}$ in the case of hard-disk interactions. For $ka_{\rm HD}=5$ (the largest value of the momentum considered here) four periods of oscillations are clearly visible. For dipolar interactions, a larger value of $ka_{\rm D}$ is required to develop a comparable oscillating pattern. For instance for $ka_{\rm D}=5$, the curve only shows a single oscillation period between $\phi=0$ and $\pi$
An oscillating behaviour is however recovered for far larger momenta, for instance the curve for $ka_{\rm D}=50$ features four well-developed minima. 


This seeming difference can be reconciled by drawing a qualitative analogy with the textbook double slit experiment where, for a fixed incident wavevector, the larger the separation between the slits, the smaller the angular separation between the fringes observed in the screen. In our case, the role of the slit separation is played by the effective radius of the repulsive potential: in the hard-disk case the wavefunction is restricted to the outer  $r>a_{\rm HD}$ region, setting the effective radius to $\bar{r}_{\rm HD}=a_{\rm HD}$. For the smoother dipolar potential the wavefunction has a finite tail in the inner region, but we can estimate the radius as the inversion point  $\bar{r}_{\rm D}$ at which the kinetic energy equals the the dipolar potential, i.e. the distance at which
\begin{align}
    \frac{k^2}{2m}=\frac{a_{\rm D}}{m \bar{r}_{\rm D}^3} \; .
\end{align}
One can expect that the differential cross section for the two potentials should display oscillations of comparable period if the two effective radii are equal $\bar{r}_{\rm D}=\bar{r}_{\rm HD}$. This implies that the incident momenta are related by 
\begin{equation}
    2ka_{\rm D}=(ka_{\rm HD})^3 \, :
    \label{eq:kaSamePeriod}
\end{equation}
the different powers appearing on either side of this equation explain why a much larger $ka_{\rm D}$ is needed to recover a $ka_{\rm HD}>1$ oscillatory pattern. Quite remarkably, this relation is approximately satisfied by the pair $k a_D=50$ and $k a_{HD}=5$: by comparing the solid cyan lines in Figs.~\ref{fig:HD_alpha05} and~\ref{fig:Dipole_alpha05}, one sees that the oscillations of these curves indeed have a very similar angular period.
The same reasoning on the effective radii $\bar{r}_{\rm HD,D}$ suggests that the angular period of the oscillations should scale as $(ka_{\rm HD})^{-1}$ and $(ka_{\rm D})^{-1/3}$, respectively. We can hint at such a dependence in the plots of Figs.~\ref{fig:HD_alpha05} and~\ref{fig:Dipole_alpha05} and a quantitative confirmation of this scaling law is provided in Appendix~\ref{sec:AppD}.



To better highlight the role of indistinguishability in determining the differential cross section, Figs.~\ref{fig:HD_alpha05} and~\ref{fig:Dipole_alpha05} also show, as dashed curves, the angular dependence of the differential scattering cross section for distinguishable impurities.
In this case, a single scattering channel contributes to the scattering cross section in each direction so the oscillating behaviour is absent and the differential cross section has a rather flat and featureless angular dependence. The small oscillations that are visible in the vicinity of the forward scattering direction $\phi=0,2\pi$ for the hard-disk case are due to diffraction effects from the sharp edges of the potential and do not have a statistical origin, as they are visible also for $\alpha=0$, as we explicitly show in of Fig.~\ref{fig:FigureAppD1} of Appendix~\ref{sec:AppD}. Still, the same figure shows that the position of the maxima and minima of these oscillations depends on the value of $\alpha$, giving an asymmetry for values different than $\alpha=0.5$.
While the divergence at $\phi=0$ due to the Aharonov-Bohm contribution \eqref{eq:ABf} is still present, the one at $\phi=\pi$ is no longer present since the forward and backward directions are no longer equivalent.
For curves displaying the same angular period in the indistinguishable case (e.g. the cyan lines in Figs.~\ref{fig:HD_alpha05} and \ref{fig:Dipole_alpha05}), the larger contrast of the fringes observed for a hard-disk potential with respect to the dipolar potential can be related to the flatter angular dependence of the distinguishable differential scattering cross section in the first case, which enhances the destructive interference.

As a next step, it is interesting to compare the two cases of bosonic and fermionic impurities (from now on, we will always consider the indistinguishable case). Fig.~\ref{fig:BosonsFermions} displays $d\sigma/d\phi$ for hard-disk bosonic and fermionic impurities with the same value $ka_{\rm HD}=1$ of the relative incident momentum and a range of values of $\nu$.
The most visible feature is a drift of the minimum towards small angles as $\nu$ is increased, with a smooth connection of the bosonic case for $\nu\to1$ and the fermionic case for $\nu\to 0$. The Bose-Fermi symmetry relations Eqs.~\eqref{eq:Bose-Fermi1}-\eqref{eq:Bose-Fermi2} also manifest in this plot.

Fig.~\ref{fig:HD_Dipolar_kaFixed} gives more details on the dependence of the differential scattering cross section on the statistical parameter $\nu$. The different panels correspond to dipolar [in a) and b)] and hard-disk [in c) and d)] interactions and to different (fixed) values of $ka_{\rm HD,D}=0.1$  [in a) and c)] and $ka_{\rm HD,D}=5$ [in b) and d)]. For standard bosons and fermions at $\nu=0$, we recover a symmetric and smooth cross section with no peaks at $\phi=0,\pi$. For bosons at intermediate values of $0<\nu<1$, the peaks at $\phi\to 0,\pi$ appear and, more interestingly, the oscillation pattern at large $k a_{\rm HD,D}$ features a global shift towards smaller angles for growing $\nu$, with again a smooth recovery to the fermionic case when $\nu\to 1$.

The linearity of this shift as a function of $\nu$ is illustrated in Fig.~\ref{fig:alphaLinear}  for both choices of interaction potential. In order to have a good contrast in the oscillations, a relatively large $k a_{\rm HD,D}$ is chosen. The deviations that are visible for the dipolar case disappear when even larger values of $k a_{\rm D}$ are chosen. For $\alpha=1$, the minimum recovers the usual location at $\phi_{\text{min}}=\pi/2$ of standard $\nu=0$ fermions. The smaller slope of the hard-disk case is a consequence of the faster angular periodicity of the oscillations that is visible when comparing the panels in Figs.~\ref{fig:HD_Dipolar_kaFixed} b) and d), both plotted for $ka_{\rm D,HD}=5$.

This simple dependence on $\nu$ of the angular interference pattern shown in Fig.~\ref{fig:alphaLinear} is a key conclusion of our study. From an experimental perspective, it provides a quantitatively accurate way to extract the fractional statistics of the quasi-holes in the FQH cloud just by detecting the oscillations in the angular dependence of the differential cross section and measuring the position of the minimum ($\phi_{\text{min}}$) in different conditions. For instance, a quantitative value of $\nu$ for a given FQH liquid can be interpolated by repeating the measurement of $\phi_{\text{min}}$ with bosonic impurities in the presence and in the absence (i.e. $\nu=0$) of the FQH liquid, keeping in mind that for $\nu=1$ the minimum is at an angle $\phi_{\text{min}}=\pi/2$, and assuming the linear dependence on $\nu=\alpha$ shown in Fig.~\ref{fig:alphaLinear}. Perhaps less challenging, a qualitative signature of the fractional statistics for $0<\nu<1$ is already offered by the asymmetry of the differential cross section for $\phi \leftrightarrow \pi-\phi$ (or, equivalently, for $\phi \leftrightarrow -\phi$ or $\phi \leftrightarrow 2\pi-\phi$), that indicates a preferential chirality in the scattering process.

From a conceptual viewpoint, the linear dependence of the fringe position on $\nu$ suggests an intuitive understanding of the underlying physical mechanism: the oscillations can be interpreted as an interference pattern for the two scattering channels contributing to the scattering in a given direction, say at an angle $\phi$. In one channel, each particle is deflected by an angle $\phi$ during the scattering process. In the other channel, each particle is deflected by $\pi+\phi$. Because of indistinguishability, the two processes have to be summed up with a relative phase $\alpha$ resulting from the sum of the intrinsic statistics $\alpha_{\text{i}}=0,1$ of the bosonic/fermionic impurities and of the fractional statistics $\nu$ of the attached quasi-holes. As it happens in generic interference experiments, e.g.  two-slit interference, a phase-shift on one of the two arms results in a rigid shift of the whole fringe pattern. This intuitive interpretation is further confirmed by the complete disappearance of the fringe pattern when distinguishable impurities are considered.

\subsection{Experimental remarks}
\label{sec:expt_scatt}
In practice, a scattering experiment will begin with the simultaneous generation of a pair of anyonic molecules at different and controlled spatial locations. The two molecules will then have to be pushed against each other at a controlled speed with suitable potentials. This can be done following one of the schemes discussed in Sec.~\ref{subsec:Expt}.

The angular dependence of their differential scattering cross section will be finally extracted by repeating the experiment many times and collecting the statistical distribution of the trajectories of the scattering products. For instance, the position of the impurities after the scattering event can be measured using absorption imaging as in Ref.~\cite{Olmos_2011}. In order to directly access the differential scattering cross section in a single shot, one may follow the route of~\cite{Kjaergaard_2004} and consider the collision between two clouds of independent impurities embedded in the FQH droplet. A possible alternative is to follow a similar strategy to that of neutron scattering in liquid helium~\cite{Yarnell_1959} and make use of a detector placed at several angular positions outside the FQH droplet. This of course has the inconvenient that the impurities may excite undesired edge modes on their way out and get their energy and momentum modified. In all cases, as we have mentioned at the beginning of Sec.~\ref{sec:TwoImpuritiesAnyPolScatt}, the analysis of the scattering experiment could be made simpler if the system parameters were chosen in such a way to give a vanishing effective charge $\mathcal{Q}$ for the anyonic molecules.

To check the actual feasibility of our proposal, it is important to estimate the maximum value of $k a_{\rm D}$ that one can realistically obtain in experiments. Combining the definition of $a_{\rm D}$ with the results of Sec.~\ref{subsec:Expt} for the maximum momentum $k_{\rm max}=\mathcal{M} v_{\rm max}$ that is compatible with the Born-Oppenheimer approach, one gets
\begin{equation}
    k_{\rm max} a_{\rm D} = \frac{\mathcal{M}}{m\tau}\frac{\Delta\omega_{-1}}{\omega_{\rm cycl}} \frac{a_{\rm D}}{\ell_B}\,.
\end{equation}
As we have seen in the previous subsection, a large value of this quantity is needed to see a well developed system of fringes in the differential scattering cross section.

As a first concrete example, we can consider the case of the magnetic interaction between two magnetic atoms, e.g. $^{166}$Er atoms with a relatively large magnetic dipole of $7\,\mu_B$~\cite{Chomaz_16}. Estimating $\mathcal{M}/m\simeq 25$, $\ell_B\simeq 1\,\mu$m, $\Delta\omega_{-1}/\omega_{\rm cycl}=0.1$, we obtain a not-so-optimistic value $k_{\rm max} a_{\rm D}\simeq 0.02$. However, a huge enhancement of the dipolar interaction is found if electric rather than magnetic interactions are used, e.g. between ground state heteronuclear diatomic molecules~\cite{Carr_2009}. For a typical dipole moment $d_{\rm E}\sim 1$~Debye, an enhancement on the order of $\approx 200$'s can be obtained, leading to a promising $k_{\rm max} a_{\rm D} \approx 5$. Thanks to the quadratic dependence of the dipolar force on the dipole moment $d_E$, a sizable further increase is achievable with specific choices of molecules which display larger dipole moments of several Debye~\cite{deiglmayr2008calculations}, e.g. 1.25~Debye for RbCs, 2.4 Debye for  NaK, up to 5.5~Debye for LiCs, and are presently under active experimental investigation in the ultracold quantum gases community~\cite{RbCs,NaK,NaK2,NaLi,NaRb,LiCs}. Note that a large value of $a_{\rm D}$ is also essential to fulfill (for a given $k a_{\rm D}$) the condition
\begin{equation}
    \frac{\Delta\omega_{-1}}{\omega_{\rm cycl}}\frac{\mathcal{M}}{m} \left(\frac{a_{\rm D}}{\ell_B}\right)^2 \gg (k a_{\rm D})^2
\end{equation}
that guarantees, according to our discussion in Sec.~\ref{subsec:Expt}, that the kinetic energy is low enough for the anyonic molecule to behave as a rigid object during the collision process.

%

\section{Conclusions}
\label{sec:Conclusions}

In this work we have shown how the quantum dynamics of impurity atoms immersed in a two-dimensional fractional quantum Hall (FQH) fluid of ultracold atoms may reveal crucial information about the fractional charge and statistics of the FQH quasi-hole excitations. Even though the discussion was carried out with a special attention to realizations in ultracold atomic gas platforms, equally promising candidates for experimental observation of the fractional charge and statistics are offered by FQH fluids of photons~\cite{Ozawa_2019,Clark_2020} or hybrid electronic-optical systems~\cite{Ravets2018,Cotlet2019}.

We considered impurities that repulsively interact with the atoms of the FQH fluid. In this case, for suitable parameters the impurities can form bound states with quasi-hole excitations, the so-called anyonic molecules.  
A rigorous Born-Oppenheimer~\cite{BO,Scherrer_2017} framework was set up to derive the effective charge and statistics of the anyonic molecules. Quite remarkably, this same formalism provides a quantitative prediction for the effective mass of the molecules, which combines the bare impurity mass with a correction due to the quasi-hole inertia.

As a main result of our work we proposed and characterized specific configurations where the fractional charge and statistics can be experimentally highlighted with state-of-the-art technology.
If a single anyonic molecule is prepared inside the FQH fluid with some initial momentum, the values of the renormalized mass and of the fractional charge can be extracted from the experimentally accessible cyclotron orbit that it describes as a free charged particle in a magnetic field. This provides direct and unambiguous information on the fractional charge of FQH quasi-holes.

In the case of two anyonic molecules, the fractional statistics of the quasi-holes provides a long-range Aharonov-Bohm-like interaction between the molecules with dramatic consequences on two-body scattering processes. For sufficiently large values of the relative incident momentum, the differential cross section displays a clear oscillatory pattern due to the interference of direct and exchange processes and the non-trivial fractional statistical phase that the quasi-holes acquire upon exchange is directly observable as a rigid shift of the angular interference pattern.

As future perspectives, we envision to extend our approach to the case of impurities binding with different numbers of quasi-holes, leading to molecules with different anyonic statistics, and to the case of a larger number of molecules forming few-body complexes with a richer structure of eigenstates determined by the interplay of the inter-impurity interaction and the fractional statistics~\cite{Wu_1984_2}. An even more intriguing development will be to extend our treatment to more subtle FQH fluids supporting non-Abelian excitations~\cite{Tong_2016} and explore the consequences of the topological degeneracy on the quantum dynamics of the non-Abelian anyonic molecules~\cite{Bonderson_2011,Zhang_2015}. 
%
\section{Acknowledgements}
\label{sec:Acknowledgements}

We acknowledge financial support from the European Union FET-Open grant ``MIR-BOSE'' (n. 737017), from the H2020-FETFLAG-2018-2020 project "PhoQuS" (n.820392), from the Provincia Autonoma di Trento, from the Q@TN initiative, and from Google via
the quantum NISQ award. Stimulating discussions with Atac Imamoglu, Gwendal F\`eve, and Giacomo Lamporesi are warmly acknowledged.

\appendix

%
\section{Mass renormalization for particles inside a magnetic field}
\label{sec:AppA}

In this first Appendix, we show how the Born-Oppenheimer (BO) approach of Ref.~\cite{Scherrer_2017} can be straightforwardly extended to include synthetic magnetic fields.

We start by writing the full action functional 
\begin{widetext}
\begin{multline}
    \mathcal{S} 
    [\varphi_\bold{R}^*,\varphi_\bold{R},\chi^*,\chi]=\bra{\psi}H-i\partial_{t}\ket{\psi}=
    \int_{t_{\text{i}}}^{t_{\text{f}}}dt \int d\bold{R} \int d\bold{r}
    \Bigg[ |\chi|^2\varphi_{\bold{R}}^*
    \left( H_{\text{BO}}+\sum_{j=1}^{N}\frac{(-i\bold{\nabla}_{\bold{R}_{j}})^2}{2M}-i\partial_{t} \right)\varphi_{\bold{R}}\\
    +|\varphi_{\bold{R}}|^2\chi^*
    \left(\sum_{j=1}^{N}\frac{(-i\bold{\nabla}_{\bold{R}_{j}}-Q\bold{A}(\bold{R}_{j}))^2}{2M} -i\partial_{t}\right)\chi
    +|\chi|^2\varphi_\bold{R}^*\sum_{j=1}^{N}
    \frac{1}{M}\frac{(-i\bold{\nabla}_{\bold{R}_{j}}-Q\bold{A}(\bold{R}_{j}))\chi}{\chi}\cdot(-i\bold{\nabla}_{j})\varphi_{\bold{R}}
    \Bigg] \; ,
\end{multline}
\end{widetext}
in terms of the atoms and impurities wavefunctions $\varphi_{\bold{R}}(t)$ and $\chi(t)$. The time dependence of the former is a finite order contribution and it is dropped when one refers to the zeroth order ground state $\varphi^{(0)}_{\bold{R}}$. As in the main text, the shorthands $\bold{R}$ and $\bold{r}$ refer to the set of impurity and atom positions $\{\bold{R}_{i}\}$ and $\{\bold{r}_{i}\}$, respectively. $\bold{A}(\bold{R}_{j})$ is the vector potential (not present in the original formulation of Ref.~\cite{Scherrer_2017}) evaluated at the position of the impurity $j$. Requiring
\begin{align}
    \frac{\delta\mathcal{S}[\varphi_\bold{R}^*,\varphi_\bold{R},\chi^*,\chi]}{\delta\varphi_{\bold{R}}^*}=0 \; , \;\;\;\;
    \frac{\delta\mathcal{S}[\varphi_\bold{R}^*,\varphi_\bold{R},\chi^*,\chi]}{\delta\chi^*}=0 \; ,
\end{align}
and using Eqs.~\eqref{eq:ExactFactorization} and~\eqref{eq:PartialNormalization} we obtain the expressions to be satisfied by the factorized wave functions
\begin{align}
\label{eq:phiExactFact}
    &[H_{\text{BO}}+U_{\text{ia}}[\varphi_{\bold{R}},\chi]-\epsilon(\bold{R},t)]\varphi_\bold{R}=i\partial_{t}\varphi_\bold{R} \; , \\
    &\left[
    \frac{(-i\bold{\nabla}_{\bold{R}_{j}}-Q\bold{A}(\bold{R}_{j})+\bold{\mathcal{A}}_j(\bold{R},t))^2}{2M}+ \epsilon(\bold{R},t)\right]\chi=i\partial_{t}\chi \; ,
\end{align}
where
\begin{align}
    &U_{\text{ia}}[\varphi_{\bold{R}},\chi]=\sum_{j=1}^{N}\frac{1}{M}\Bigg[
    \frac{(-i\bold{\nabla}_{\bold{R}_{j}}-\bold{\mathcal{A}}_{j}(\bold{R},t))^2}{2}+\nonumber\\
    &\left(\frac{(-i\bold{\nabla}_{\bold{R}_{j}}-Q\bold{A}(\bold{R}_{j}))\chi}{\chi}+\bold{\mathcal{A}}_j(\bold{R},t)\right)
    \cdot(-i\bold{\nabla}_{\bold{R}_{j}}-\bold{\mathcal{A}}_{j}(\bold{R},t))
    \Bigg] 
\end{align}
is the impurity-bath coupling operator,
\begin{align}
    \bold{\mathcal{A}}_j(\bold{R},t)=-i \bra{\varphi_{\bold{R}}(t)}\bold{\nabla}_{\bold{R}_{j}}\ket{\varphi_{\bold{R}}(t)}
\end{align}
is the Berry connection arising from the parametric dependence of the bath wave function on the position of the impurities and
\begin{align}
\label{eq:tdes}
    \epsilon(\bold{R},t)=\bra{\varphi_{\bold{R}}(t)}H_{\text{BO}}+U_{\text{ia}}-i\partial_{t}
    \ket{\varphi_{\bold{R}}(t)} 
\end{align}
is the BO potential energy surface experienced by the moving impurities, which mediates the exact coupling between fast and slow degrees of freedom.

We see that the complete expressions in Eqs.~\eqref{eq:phiExactFact}-\eqref{eq:tdes} do not include any additional first order modification beyond those calculated in Ref.~\cite{Scherrer_2017}. Therefore, the renormalized mass of the  molecule keeps the same form as in the case without vector potential $\bold{A}(\bold{R}_{j})$, which we show in Eqs.~\eqref{eq:MassRenormalization}-\eqref{eq:phi1}.

%
\section{Calculation of the $\tau$ coefficient}
\label{sec:AppB}
\begin{figure}
    \centering
    \includegraphics[width=0.5\textwidth]{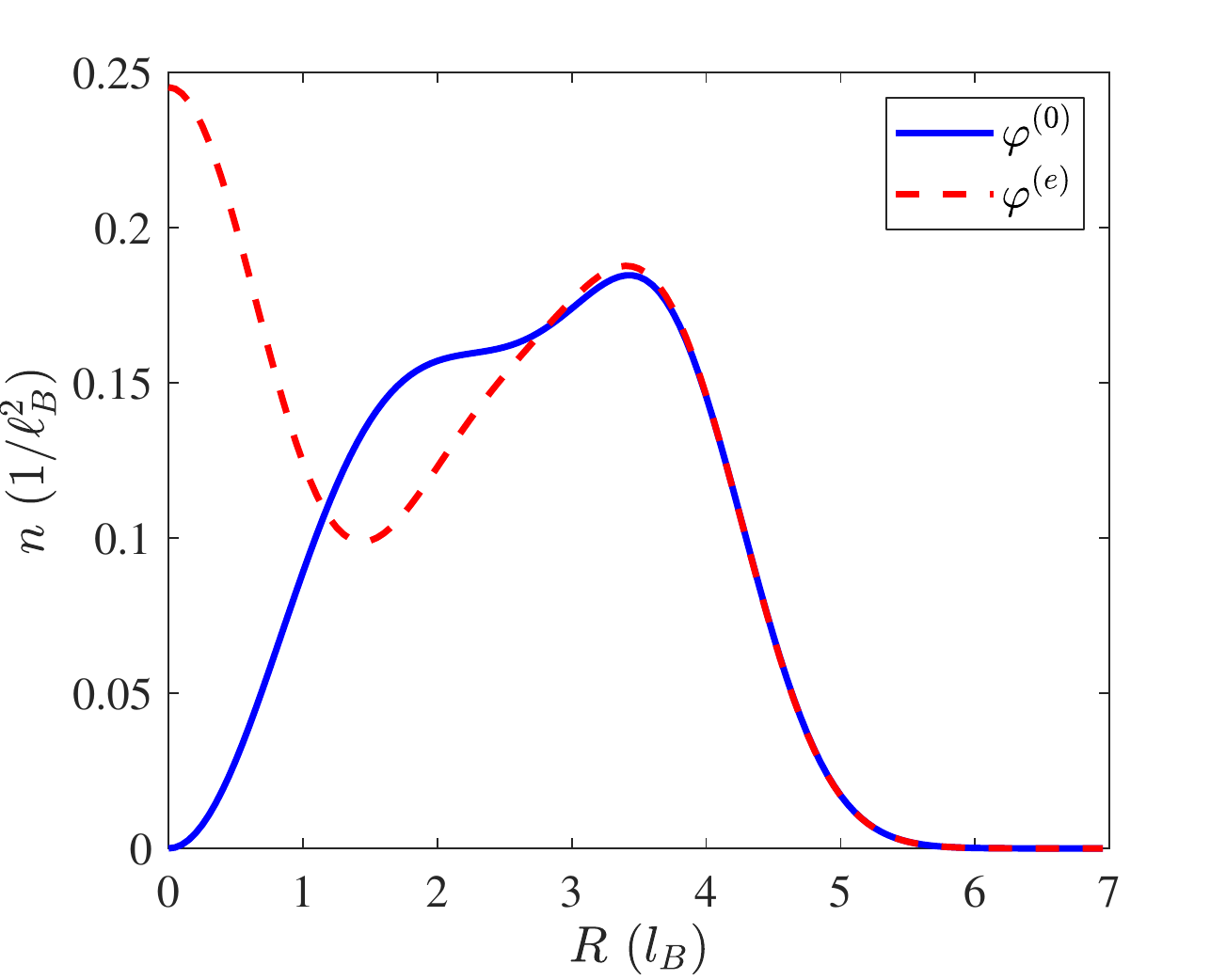}
    \caption{Density of the ground state $\varphi^{(0)}$ and the first excited state $\varphi^{(\text{e})}$ as function of the distance to the quasi-hole located at $R=0$ in the $\varphi^{(0)}$ case. Both curves were calculated for $n=10$ particles in the FQH bath and are converged w.r.t. $n$.}
    \label{fig:FigureAppB}
\end{figure}

In this Appendix, we report the calculations leading to the numerical value $\tau \simeq 0.7$ for the coefficient $\tau$ introduced in Eq.~\eqref{eq:tau} and used to quantitatively estimate the mass correction $\Delta M$ in Eq.\eqref{eq:phi_e}.

We denote with $\ket{\varphi^{(0)}_Z}$ the quantum state of the atomic fluid with one quasi-hole at the position $\bold{R}$ corresponding to the complex variable $Z=X-iY$. We then expand the $Z$-dependent state vector around $Z=0$ as 
%
\begin{multline}
    \ket{\varphi^{(0)}_Z}
    = \ket{\varphi^{(0)}_{Z=0}}
    +Z\left(\partial_{Z}\ket{\varphi^{(0)}_{Z}}\right)
    +Z^{*}\left(\partial_{Z^{*}}\ket{\varphi^{(0)}_{Z}}\right)\nonumber\\
    +\frac{Z^2}{2}\left(\partial^{2}_{Z}\ket{\varphi^{(0)}_{Z}}\right)
    +\frac{Z^{*2}}{2}\left(\partial^{2}_{Z^{*}}\ket{\varphi^{(0)}_{Z}}\right)\nonumber\\
    +|Z|^2\,\left(\partial_{Z}\partial_{Z^{*}}\ket{\varphi^{(0)}_{Z}}\right) + \mathcal{O}(Z^3)\; .
\end{multline}
where all derivatives are evaluated at $Z=0$.

We consider the atomic density operator $\hat{n}(r)$ evaluated at position $r$, indicated by the complex variable $z$. Taking advantage of the fact that
\begin{align}
&\hat{n}(z=Z)\ket{\varphi^{(0)}_{Z}}=0\; , \\ &\left(\partial_{Z^*}\ket{\varphi^{(0)}_{Z}}\right)_{Z=0}=0\; , \\
&\partial_{Z}\ket{\varphi^{(0)}_{Z}}=\frac{\tau}{\ell_B}\ket{\varphi^{(e)}_{Z}}\, , 
\end{align}
we get that
%
%
\begin{multline}
    \bra{\varphi^{(0)}_{Z}}\hat{n}(z=0)\ket{\varphi^{(0)}_{Z}}
    \simeq \\ \simeq |Z|^2\left(\partial_{Z^*}\bra{\varphi^{(0)}_{Z}}\right) \hat{n}(z=0)\left(\partial_{Z}\ket{\varphi^{(0)}_{Z}}\right) = \\ 
    =|Z|^2\frac{\tau^2}{\ell^2_{\text{B}}}\bra{\varphi^{(\text{e})}_{Z}}\hat{n}(z=0)\ket{(\varphi^{(\text{e})}_{Z}}\; .
\end{multline}
For a quasi-hole living in the bulk of the FQH droplet we can take advantage of the local homogeneity of the atomic fluid to write
\begin{multline}
 \bra{\varphi^{(0)}_{Z=0}}\hat{n}(z)\ket{\varphi^{(0)}_{Z=0}}=
     \bra{\varphi^{(0)}_{Z=z}}\hat{n}(z=0)\ket{\varphi^{(0)}_{Z=z}} \simeq \\ \simeq |z|^2\frac{\tau^2}{\ell^2_{\text{B}}}\bra{\varphi^{(\text{e})}_{Z}}\hat{n}(z=0)\ket{\varphi^{(\text{e})}_{Z}}= \\ =|z|^2\frac{\tau^2}{\ell^2_{\text{B}}}\bra{\varphi^{(\text{e})}_{Z=0}}\hat{n}(z)\ket{\varphi^{(\text{e})}_{Z=0}}\; . \label{eq:tau_dens}
\end{multline}
This expression relates the value of $\tau$ to the (normalized) radial density distribution $n_0(r)$ of the ground state $\varphi^{(0)}_{\bold{R}}$ 
and the distribution $n_{\rm e}(r)$ in the first excited  $\varphi^{(\text{e})}_{\bold{R}}$ 
state. These two distributions can be numerically calculated using the expansion of $\varphi^{(\text{0})}_{\bold{R}}$ and $\varphi^{(\text{e})}_{\bold{R}}$ in terms of Jack polynomials developed in Ref.~\cite{Macaluso_2018}. The results for a mesoscopic cloud of $n=10$ particles are displayed in Fig.~\ref{fig:FigureAppB} and we have made sure that our results for the inner part of the profile are converged with respect to  the number of particles $n$.
In particular, we see in the figure that $n_0(\ell_{\rm B}/2) 
\simeq 0.028/\ell^{2}_{\text{B}}$, while $n_{\rm e}(0)
\simeq 0.245/\ell^{2}_{\text{B}}$. Substituting these values into Eq.~\eqref{eq:phi_e} one obtains $\tau\simeq 0.7$.

%
\section{Energy gap and impurity-atom interaction}
\label{sec:AppC}
\begin{figure}
    \centering
    \includegraphics[width=0.5\textwidth]{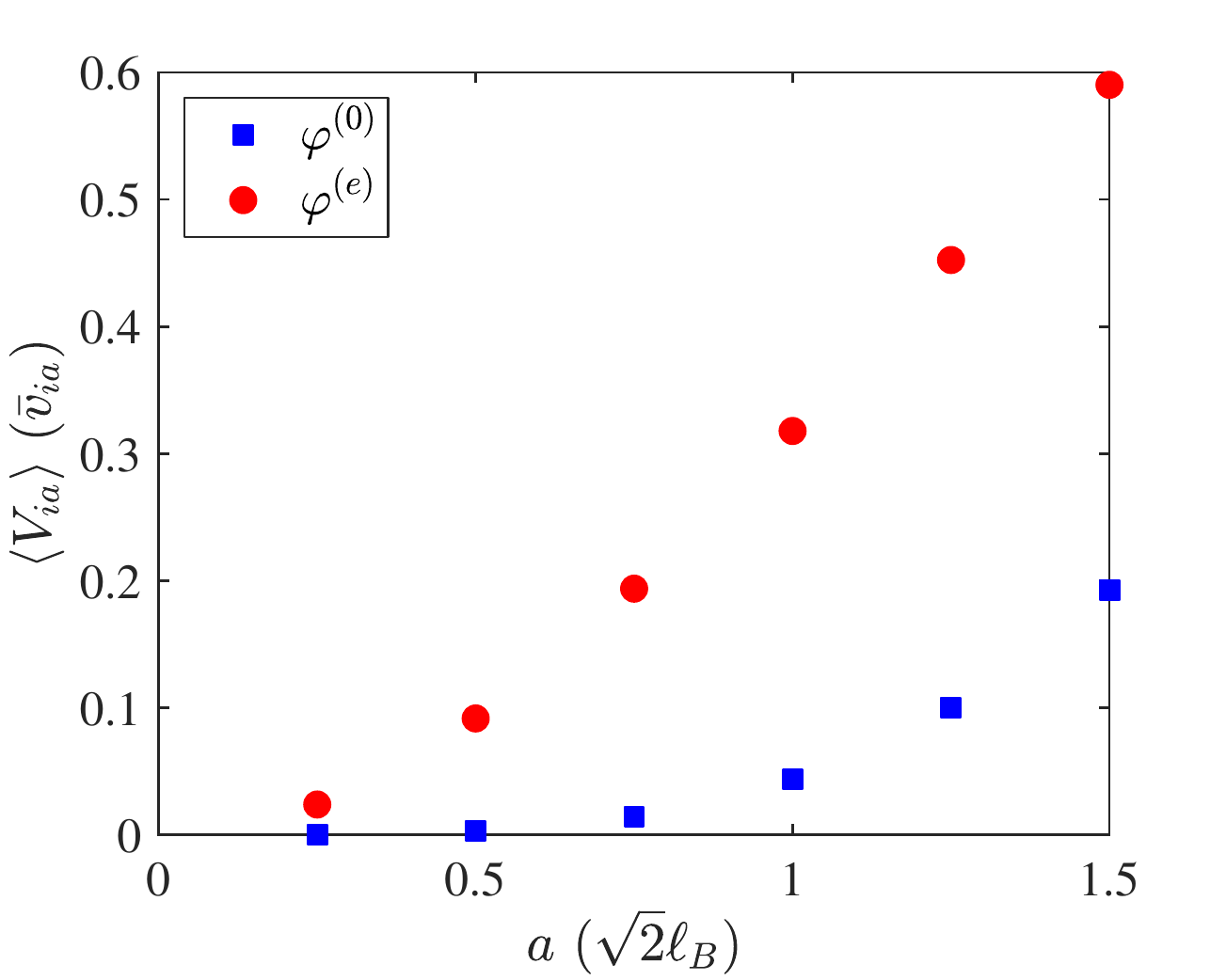}
    \caption{Expectation value of the step-like interaction between impurities and atoms $V_{ia}$ in the ground state $\varphi^{(0)}$ (blue squares) and in the lowest excited state $\varphi^{(e)}$ (red circles) in units of $\bar{v}_{\text{ia}}$ as a function of its step radius $a$ (in units of $\ell_{\text{B}}$).}
    \label{fig:FigureAppC}
\end{figure}

In this third Appendix, we offer quantitative evidence of the binding of impurities to FQH quasi-holes in the presence of a repulsive interaction between the impurity and the atoms.

To this purpose, we calculate the dependence of the energy difference between the ground state $\varphi^{(0)}$ displaying a quasi-hole located at the impurity's position (taken as the origin, i.e. $\bold{R}=0$) and the lowest excited state $\varphi^{(e)}$ with the quasi-hole orbiting around the impurity with $\Delta L=-1$ w.r.t. the impurity-atom interaction. For simplicity, we employ a step-like interaction potential between the impurity and the atoms of the FQH fluid of the form
\begin{align}
    V_{\text{ia}}=\iac{\bar{ v}}_{\text{ia}}\sum^{n}_{i=1}\Theta (a-|\bold{r}_{i}|) \; ,
\label{eq:Via}
\end{align}
where $n$ is the number of atoms, $\bar{v}_{\text{ia}}>0$ and $a$ is the step radius.

We first compute the states of interest using a Jack polynomials expansion~\cite{Macaluso_2018} and then calculate the expectation values of the atom-impurity potentials in the ground and lowest excited states for different values of the step radius $a$. We have checked that the density profiles of the relevant many-body states in the vicinity of the impurity (i.e. for distances $\leq a$) do not depend on the total number $n$ of atoms in the FQH state. 

The expectation values in the ground state and in the lowest excited state for several values of $a$ are shown  in Fig.~\ref{fig:FigureAppC} as blue squares and red circles, respectively. While the ground state is almost insensitive to the presence of the impurity as long as the radius $a$ of the interaction potential is much smaller than the spatial extension of the density depletion of the quasi-hole (on the order of the magnetic length $\ell_{\rm B}$, as shown by the blue curve in Fig.~\ref{fig:FigureAppB}), the density in the excited state always has a significant overlap with the impurity (as shown in the red curve in Fig.~\ref{fig:FigureAppB}), which gives a sizable energy shift of this state that grows quadratically with $a$ and has a finite limit in the contact limit $a\to 0 $ at a constant $\bar{v}_{\rm ia} a^2$.

The significant resulting energy gap $\Delta\omega_{-1}$ between the two states leads then to an efficient binding of the QH to the impurity. As the step radius or the potential strength are increased, its magnitude increases, thus reinforcing the rigidity of the impurity-quasihole molecule and reducing the importance of the Born-Oppenheimer mass correction $\Delta M$ in Eq. \eqref{eq:phi_e}. 

%
\section{On the oscillations of the differential scattering cross section}
\label{sec:AppD}
\begin{figure}
    \centering
    \includegraphics[width=0.5\textwidth]{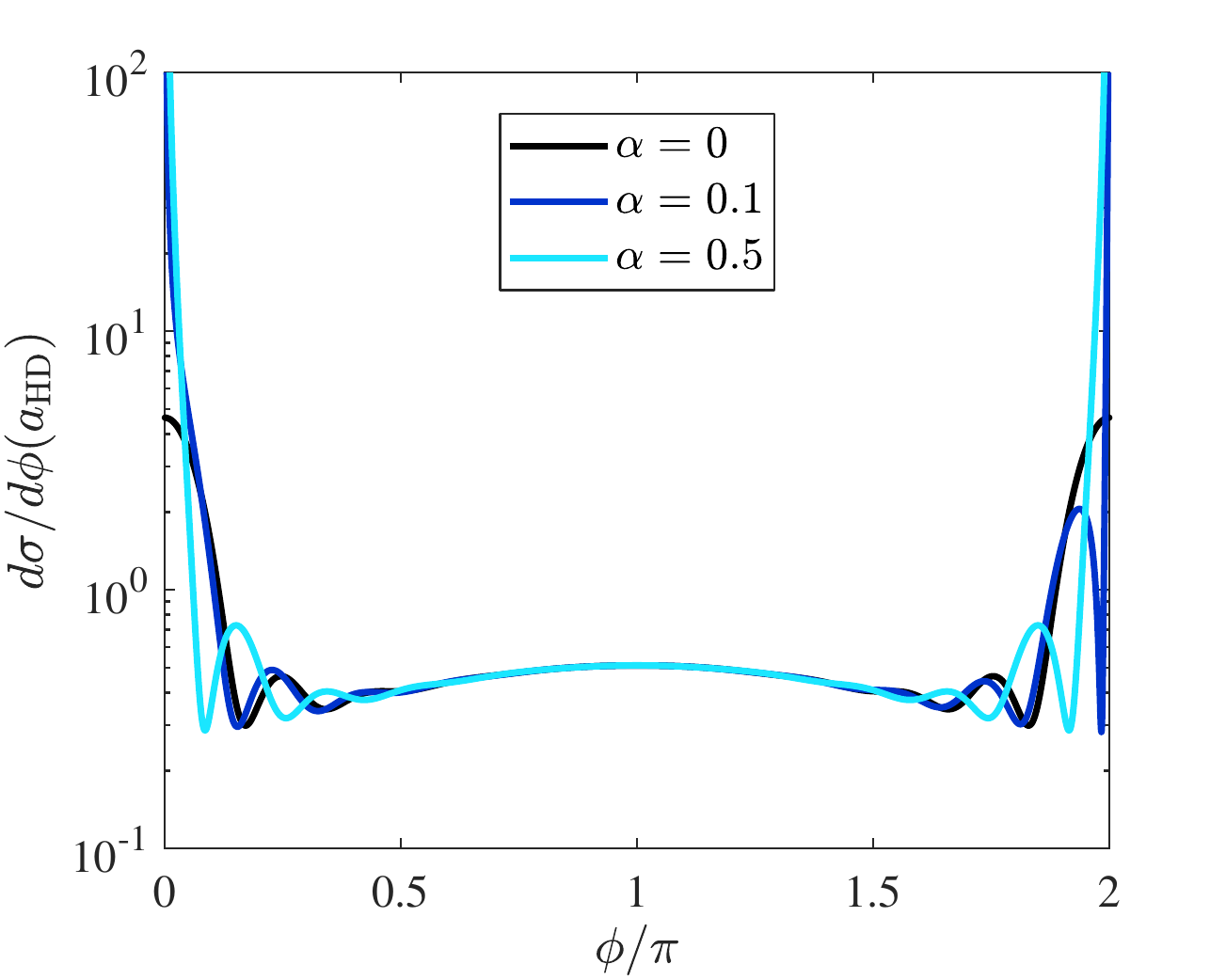}
    \caption{Differential scattering cross section $d\sigma/d\phi$ in units of the hard-core radius $a_{\rm HD}$ as function of the scattering angle $\phi$ for a hard-disk interaction potential $V_{ii}$ between distinguishable impurities with a relative incident momentum $ka_{\rm HD}=5$, and for three values of the statistical parameter $\alpha$.}
    \label{fig:FigureAppD1}
\end{figure}
\begin{figure*}
    \centering
    \includegraphics[width=\textwidth]{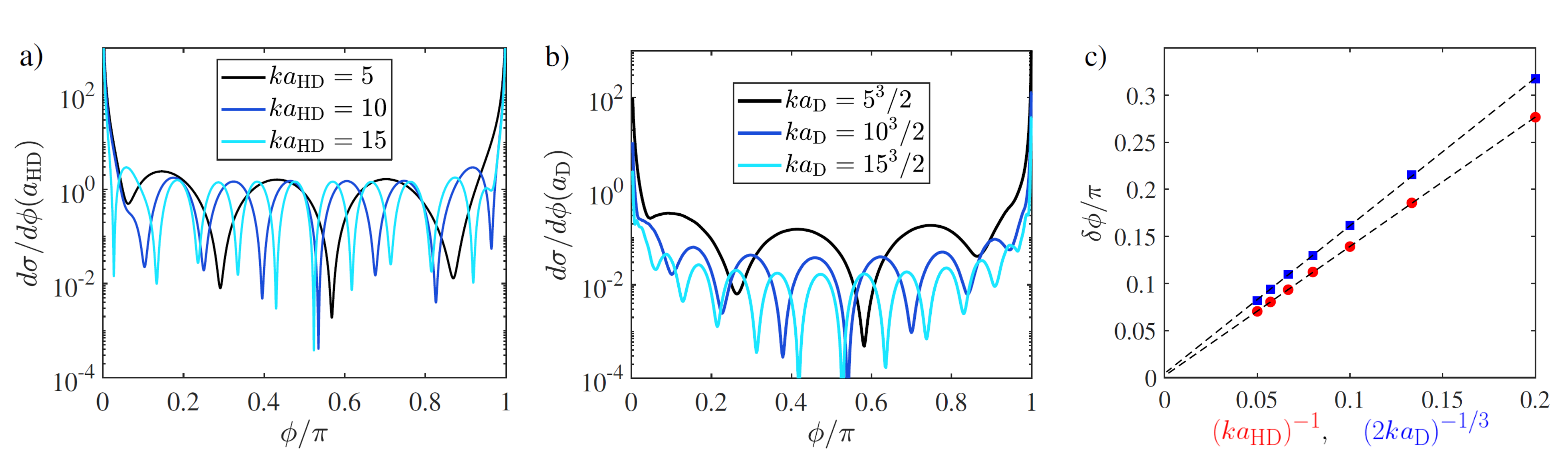}
    \caption{Differential scattering cross section $d\sigma/d\phi$ as function of the scattering angle $\phi$ for indistinguishable molecules with statistical parameter $\alpha=0.5$ and several values of the relative incident momentum $ka_{\rm HD,D}$ in the hard-disk \textbf{a)} and dipolar \textbf{b)} interaction cases. The curves plotted with the same color in these two panels satisfy the relation Eq.~\eqref{eq:kaSamePeriod} and are thus expected to have the same angular oscillation period. Panel \textbf{c)} shows the oscillation period $\delta\phi$ (calculated as the difference between the angular position of the two central minima) as a function of $(ka_{\rm HD})^{-1}$ for hard-disk impurities (red circles) and as a function of $(2ka_{\rm D})^{-1/3}$ for dipolar ones (blue squares). The dashed black lines are the respectives linear fits.}
    \label{fig:FigureAppD2}
\end{figure*}

In this last Appendix we provide additional quantitative evidence on the oscillatory behavior of the differential scattering cross section described in Sec.~\ref{sec:Numerics}.

Fig.~\ref{fig:FigureAppD1} shows the angular dependence of the differential scattering cross section for distinguishable bosons ($\alpha=0$; black curve) and for distinguishable anyonic molecules with $\alpha=0.1$ (dark blue curve) and $\alpha=0.5$ (cyan curve) interacting via a hard-disk potential for a fixed relative incident momentum $ka_{\rm HD}=5$. The latter case was previously displayed as the dashed line in Fig.~\ref{fig:HD_alpha05}. 

The two curves present a very similar and featureless behavior for angles $0.5\pi<\phi< 1.5\pi$, confirming the crucial role of indistinguishability and of interference between the direct and exchange scattering channels.

The oscillations displayed in the vicinity of the forward scattering direction for $\phi=0,2\pi$ are instead due to diffraction effects around the hard-disk potential and do not have any statistical origin. They are in fact present in the three curves with a comparable angular period. Of course, the anyonic curves show the expected divergence for $\phi \rightarrow 0$, which is not present in the bosonic case. Apart from this, a finite $\alpha$ introduces a phase shift of the diffraction pattern and, for values $\alpha\neq 0.5$, an asymmetry between the oscillations to the left and to the right of the forward direction (see the dark blue curve). 

In Fig.~\ref{fig:FigureAppD2} we compare the angular dependence of the differential scattering cross section for the hard-disk (left panel) and dipolar (central panel) cases for several parameter choices that pairwise satisfy the relation Eq.~\eqref{eq:kaSamePeriod} between the incident wavevectors. As expected, the oscillation periods are almost identical within each pair (represented by the same color in Fig.~\ref{fig:FigureAppD2}). As a further illustration, panel c) demonstrates the linear dependence of the oscillation period w.r.t. $(ka_{\rm HD})^{-1}$ and $(2ka_{\rm D})^{-1/3}$ in each case, which confirms our intuitive interpretation in terms of the textbook double slit experiment.

\bibliography{scattering_anyonic_polarons}

\end{document}